\documentclass[aip, pof,graphicx]{revtex4-2}

\usepackage{graphicx}
\usepackage{dcolumn}
\usepackage{bm}
\usepackage[mathlines]{lineno}
\usepackage[utf8]{inputenc}
\usepackage[T1]{fontenc}
\usepackage{mathptmx}
\usepackage{color}
\usepackage{longtable}
\usepackage{amssymb}

\draft 

\begin{document}


\title[Revealing the intricate dune-dune interactions of bidisperse barchans]{Revealing the intricate dune-dune interactions of bidisperse barchans}



\author{Willian R. Assis}
\affiliation{School of Mechanical Engineering, UNICAMP - University of Campinas,\\
	Rua Mendeleyev, 200, Campinas, SP, Brazil
}%

\author{Fernando David C\'u\~nez}
\affiliation{Department of Earth and Environmental Sciences, University of Rochester,\\
	Rochester, NY 14627, USA}

\author{Erick M. Franklin*}%
 \email{erick.franklin@unicamp.br}
 \thanks{*Corresponding author}
\affiliation{School of Mechanical Engineering, UNICAMP - University of Campinas,\\
Rua Mendeleyev, 200, Campinas, SP, Brazil
}%


\date{\today}


\begin{abstract}
Three dimensional dunes of crescentic shape, called barchans, are commonly found on Earth and other planetary environments. In the great majority of cases, barchans are organized in large fields in which corridors of size-selected barchans are observed, and where barchan-barchan interactions play an important role in size regulation. Previous studies shed light on the interactions between barchans by making use of monodisperse particles, but dunes in nature consist, however, of polydisperse grains. In this paper, we investigate the binary interactions of barchans consisting of (i) bidisperse mixtures of grains and (ii) different monodisperse grains (one type for each barchan). We performed experiments in a water channel where grains of different sizes were poured inside forming two barchans that interacted with each other while filmed by a camera, and we obtained their morphology from image processing. We observed that a transient stripe appears over the dunes in cases of bidisperse mixtures, that interaction patterns vary with concentrations, and that different interactions exist when each barchan consists of different monodisperse grains. Interestingly, we found the conditions for a collision in which the upstream barchan is larger than the downstream one, and we propose a timescale for the interactions of both monodisperse and bidisperse barchans. Our results represent a new step toward understanding complex barchanoid structures found on Earth, Mars and other celestial bodies.
\end{abstract}


\maketitle 


\section{INTRODUCTION}
\label{sec:intro}

Under one-directional fluid flow and limited amount of sand, three dimensional dunes of crescentic shape, called barchans, consistently grow \cite{Bagnold_1, Herrmann_Sauermann, Hersen_3}, being commonly found on Earth, Mars, other celestial bodies \cite{Elbelrhiti, Claudin_Andreotti, Parteli2}. In the great majority of cases, barchans are organized in large fields in which corridors of size-selected barchans are observed, and where barchan-barchan interactions play an important role in size regulation \cite{Hersen_2, Hersen_5, Kocurek, Genois, Genois2, Assis, Assis2}. Given the ubiquitous nature of barchans, understanding how their shape is formed, how they self organize in regular fields, and how they are affected by the bed granulometry are of paramount importance to deduce the past and predict the future of barchans on Earth and other planetary environments.

Several studies investigated the interactions between barchans, shedding light on certain aspects of barchan-barchan interations, but leaving many others, however, poorly understood. Among the drawbacks of previous studies, field measurements were very limited in time, since aeolian barchans take decades to complete interact with each other \cite{Bagnold_1, Hersen_1}, and previous experimental and numerical studies made use of monodisperse particles, whereas dunes in nature consist of polydisperse grains. Therefore, one way to overcome those drawbacks is by carrying out experiments with polydisperse grains under water, the subaqueous barchans being much faster and smaller than the aeolian dunes (the former having time and length scales of the order of minutes and centimeters \cite{Hersen_1, Franklin_2, Franklin_8, Alvarez}).

Field measurements of aeolian barchan-barchan interactions are of course important, consisting in a direct observation of nature, and past studies \cite{Norris, Gay, Vermeesch, Elbelrhiti2, Hugenholtz} showed that the collisions of barchans regulate their size and generate different barchanoid forms. However, the time series are frequently incomplete (given the large timescales involved), and, therefore, numerical and experimental investigations have been conducted in parallel with field experiments.

The numerical studies made use of continuum \cite{Schwammle2, Duran2, Zhou2} and discrete \cite{Katsuki} models to compute the evolution of a bed surface into a dune field, most of them incorporating rules for barchan-barchan interations \cite{Lima, Partelli5, Katsuki2, Duran3}. In addition to these techniques, Duran et al. \cite{Duran3} and G\'enois et al. \cite{Genois2} proposed an agent-based model that makes use of sand flux balances and elementary rules for barchan collisions, and Bo and Zheng \cite{Bo} used a scale-coupled model \cite{Zheng} to obtain the probability of occurrence of different types of barchan-barchan collisions. Those numerical investigations showed that barchan-barchan collisions lead to corridors of size-selected barchans, pointing toward homogeneous fields. However, model simplifications prevented them from reproducing correctly all barchan-barchan interactions, and, in addition, there is a lack of numerical studies at the grain scale.

The experiments were carried out almost exclusively in water tanks and channels. Some studies measured the flow disturbances caused by an upstream barchan upon a downstream one, such as done by Bristow et al. \cite{Bristow, Bristow2, Bristow3}, who found, among other findings, that turbulence levels increase on the stoss surface of the downstream dune, enhancing erosion over the downstream dune. Other investigations measured the evolution of two interacting bedforms, identifying different interaction patterns \cite{Endo2, Hersen_5, Bacik, Assis} and mass exchanges at the grain scale \cite{Assis2}. Endo et al. \cite{Endo2} investigated collisions of aligned barchans by varying their mass ratio while maintaining fixed the water flow rate, initial conditions and grain types, and they found three collision patterns which were called merging, exchange and fragmentation-chasing by Assis and Franklin \cite{Assis} (and explained next). Hersen and Douady \cite{Hersen_5} investigated the collisions of off-centered barchans by varying their transverse distances while keeping the other parameters fixed, and showed that collisions produce smaller barchans, regulating thus their size when in a barchan field. The experiments of Bacik et al. \cite{Bacik} were devoted to the interaction over long times between a pair of two-dimensional dunes in a circular channel, and they found that turbulent structures of the disturbed flow prevent dune collisions by inducing dune-dune repulsion. Recently, Assis and Franklin \cite{Assis, Assis2} inquired further into the binary interactions of subaqueous barchans by conducting experiments in both aligned and off-centered configurations where  the water flow rates, grain types (diameter, density and roundness), pile masses, longitudinal and transverse distances, and initial conditions were varied, and measurements were made at the bedform and grain scales. They found five interaction patterns for both aligned and off-centered configurations, proposed classification maps, measured the trajectories of individual grains during barchan-barchan interactions, and found the typical lengths and velocities of grains, the mass exchanged between barchans, and a diffusive length for some collisions.

From the previous works, the most comprehensive classification of barchan-barchan interactions is the one presented in Ref. \cite{Assis}, which identifies: (i) chasing, when collision does not occur, the upstream barchan not reaching the downstream one; (ii) merging, when collision occurs and the dunes merge; (iii) exchange, when, once collision takes place, a small barchan is ejected; (iv) fragmentation-chasing, when the downstream dune splits without collision taking place and the downstream bedforms outrun the upstream one; and (v) fragmentation-exchange, when fragmentation initiates, collision takes place, and a small barchan is ejected. The question that persists is if the same patterns and classification maps proposed by Assis and Franklin \cite{Assis} remain valid for aeolian and Martian barchans (and also other planetary environments), and polydisperse dunes. While proving the validity for aeolian and Martian barchans is hindered by their large timescales, that for polydisperse barchans, on the other hand, can be investigated in the subaqueous case. 

Concerning dunes of polydisperse grains, Alvarez et al. \cite{Alvarez5} investigated experimentally the growth of subaqueous single barchans consisting of bidisperse grains. In their experiments, single granular piles consisting of bidisperse mixtures in terms of grain sizes and/or densities were developed into barchan dunes, and they found that denser, smaller, and smaller and less dense grains tend to accumulate over the barchan surface. They also found that a transient stripe transverse to the flow direction appears just upstream the crest of the initial bedform and migrates toward its leading edge until disappearing, that that line separates a downstream region where segregation is complete from the upstream region where segregation is still occurring, and that the final barchan morphology is roughly the same as that of monodisperse barchans. Finally, they proposed that segregation patterns result from a competition between fluid entrainment and easiness of rolling, and showed that grains segregate with a diffusion-like mechanism.

In this paper, we investigate the binary interactions of barchans when grains of two different sizes are involved. For that, we inquired into two specific cases: (i) each bedform consisting of bidisperse mixtures; (ii) each bedform consisting of a given, but different between them, grain type (two-species monodisperse barchans). The experiments were conducted in a water channel where grains were poured inside, forming two conical piles that were afterward deformed by the water flow into barchans that interacted with each other. The evolution and interactions of bedforms were recorded by a conventional camera and their morphology was obtained from image processing. We observe that a transient stripe appears over the dunes in cases of bidisperse mixtures (just as happens for single bidisperse barchans \cite{Alvarez5}), that interaction patterns vary with concentrations, and that different interactions exist when each barchan consists of monodisperse grains of different kind (two-species monodisperse barchans), including collisions in which the upstream barchan is larger than the downstream one. Finally, we propose a timescale for the interactions of both monodisperse (one-species) and bidisperse (cases i and ii) barchans. Our results represent a new step toward understanding complex barchanoid structures found on Earth, Mars and other celestial bodies.

\section{EXPERIMENTAL SETUP}
\label{sec:methods}

\begin{figure}[h!]
	\begin{center}
		\includegraphics[width=.95\linewidth]{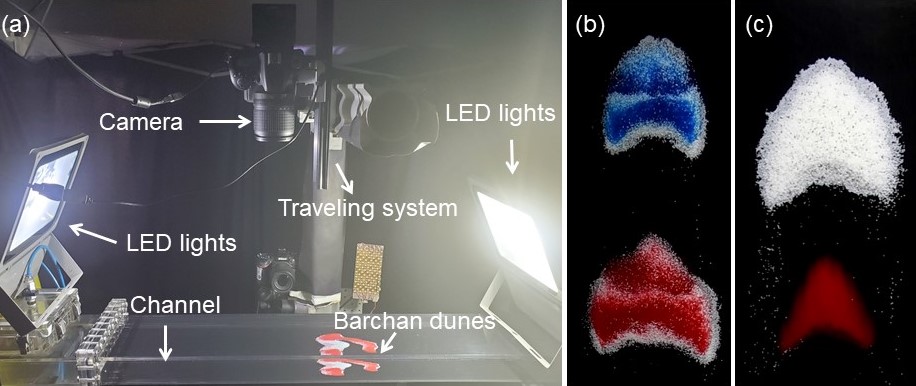}\\
	\end{center}
	\caption{(a) Photograph of test section showing the dunes, camera and LED lights; (b) top-view image of two interacting bidisperse barchans (case $e$ of Tab. \ref{tab1}); (c) top-view image of two interacting monodisperse barchans of different granulometry (case $o$ of Tab. \ref{tab1}, of a larger barchan reaching a smaller one); In figures (b) and (c) the flow is from top to bottom.}
	\label{fig:setup}
\end{figure}

The experimental device consisted basically of a water tank, centrifugal pumps, a flow straightener, a 5-m-long closed-conduit channel, a settling tank, and a return line, so that we imposed a pressure-driven water flow in closed loop. The channel had a rectangular cross section 160 mm wide by 2$\delta$ = 50 mm high, was made of transparent material, and consisted of a 3-m-long entrance section (corresponding to 40 hydraulic diameters), a 1-m-long test section, and a 1-m-long section connecting the test section to the channel exit. With the channel filled with water in still conditions, controlled grains were poured inside in order to form two aligned piles consisting of (i) bidisperse mixtures; (ii) different monodisperse grains (two-species, each pile consisting of one single species). Afterward, a specified water flow was imposed, deforming each pile into a barchan dune that interacted with each other while a camera recorded top view images of the bedforms. No influx of grains coming from regions upstream the test section was imposed and, therefore, the entire system decreased in mass along time. Using the same terminology of previous works \cite{Assis, Assis2}, we call \textit{impact barchan} the one that was initially upstream and \textit{target barchan} the one that was initially downstream. Figure \ref{fig:setup}(a) shows a photograph of the test section, and Figs. \ref{fig:setup}(b) and \ref{fig:setup}(c) top-view images of two interacting barchans consisting each of bidisperse and monodisperse grains, respectively. The layout of the experimental device is shown in the supplementary material.

In our tests, we used tap water at temperatures within 22 and 28 $^{\circ}$C and round glass beads ($\rho_s$ = 2500 kg/m$^3$) with diameters 0.15 mm $\leq$ $d_{s1}$ $\leq$ 0.25 mm and 0.40 mm $\leq$ $d_{s2}$ $\leq$ 0.60 mm, which we call species 1 and 2, respectively. We consider in our computations the mean values $d_1$ = 0.2 mm and $d_2$ = 0.5 mm of $d_{s1}$ and $d_{s2}$, respectively, and we used grains of different colors (white, red and blue) in order to track the different species along images (see the supplementary material for microscopy images of the used grains). We varied the concentration of each grain type ($\phi_1$ and $\phi_2$) between 0 and 1, the mass ratio of the piles (initial mass of the impact barchan $m_i$ divided by that of the target one $m_t$) between 0.02 and 4, and the water velocities within 0.278 m/s $\leq$ $U$ $\leq$ 0.347 ms, where $U$ is the cross-sectional mean velocity of water. These values correspond to Reynolds numbers based on the channel height, Re = $\rho U 2\delta /\mu$, within $1.39$ $\times$ $10^4$ and $1.74$ $\times$ $10^4 $, where $\mu$ is the dynamic viscosity and $\rho$ the density of the fluid. We computed shear velocities of the undisturbed water flow over the channel walls $u_*$ from measurements with a two-dimensional two-component particle image velocimetry (2D2C-PIV) device, and we use $u_*$ as a reference value for fluid shearing even when bedforms are present in the channel. Our measurements show values within 0.0159 and 0.0193 m/s and that $u_*$ follows the Blasius correlation \cite{Schlichting_1}. With those values, the Shields number $\theta = (\rho u_*^2)/((\rho_s - \rho )gd)$ varied within 0.034 and 0.127 (where $g$ is the acceleration of gravity). Table \ref{tab1} summarizes the tested conditions, and complete tables with all the parameters (in dimensional form) are available on an open repository \cite{Supplemental2}.

A digital camera with a lens of 18-140 mm focal distance and F2.8 maximum aperture was mounted on a traveling system in order to have a top view of the bedforms, and lamps of light-emitting diode (LED) were used as light source. The camera was of complementary metal-oxide-semiconductor (CMOS) type with a maximum resolution of 1920 px $\times$ 1080 px at 60 Hz, and the region of interest (ROI) was set between 1311 px $\times$ 451 px and 1920 px $\times$ 771 px, for fields of view varying within 354 mm $\times$ 122 mm and 507 mm $\times$ 122 mm. The acquired images were afterward processed by numerical scripts that identified and tracked bedforms and patterns, and were based on Ref. \cite{Crocker}. Movies showing collisions of bidisperse barchans are available in the supplementary material and on an open repository \cite{Supplemental2}.

\section{RESULTS}
\label{sec:results}

\subsection{Bidisperse piles}
\label{sec:results_bidisp}

\begin{table}[!ht]	
	\begin{center}
		\begin{tabular}{c c c c c c c c c c c c c c}
			\hline\hline
			Case & $\phi_{1t}$ & $\phi_{2t}$ & $\phi_{1i}$ & $\phi_{2i}$ & $D_t$ & $D_i$ & $\Delta x_d$ & Re & $u_*$ & $m_i/m_t$ & $t_s$ & $t_c$ & \textit{Pat}\\ 
			$\cdots$  & $\cdots$ & $\cdots$ & $\cdots$ & $\cdots$ & mm & mm & mm & $\cdots$ & mm/s & $\cdots$ & s & $\cdots$ & $\cdots$ \\\hline
			a & 0.5 & 0.5 & 0.5 & 0.5 & 63 & 18 & 18 & 1.56 $\times$ 10$^4$ & 0.0176 & 0.02 & 230 & 0.1 & M\\
			b & 0.5 & 0.5 & 0.5 & 0.5 & 61 & 24 & 23 & 1.56 $\times$ 10$^4$ & 0.0176 & 0.05 & 450 & 0.2 & E\\
			c & 0.5 & 0.5 & 0.5 & 0.5 & 56 & 27 & 20 & 1.56 $\times$ 10$^4$ & 0.0176 & 0.11 & 499 & 0.6 & FE\\
			d & 0.5 & 0.5 & 0.5 & 0.5 & 43 & 30 & 30 & 1.56 $\times$ 10$^4$ & 0.0176 & 0.43 & 1537 & $\infty$ & FC\\
			e & 0.5 & 0.5 & 0.5 & 0.5 & 52 & 48 & 39 & 1.56 $\times$ 10$^4$ & 0.0176 & 0.67 & 10637 & $\infty$ & C\\
			f & 0.8 & 0.2 & 0.8 & 0.2 & 56 & 22 & 21 & 1.56 $\times$ 10$^4$ & 0.0176 & 0.05 & 482 & 0.1 & E\\
			g & 0.2 & 0.8 & 0.2 & 0.8 & 60 & 25 & 21 & 1.56 $\times$ 10$^4$ & 0.0176 & 0.05 & 318 & 0.2 & M\\
			h & 0.8 & 0.2 & 0.2 & 0.8 & 54 & 21 & 21 & 1.56 $\times$ 10$^4$ & 0.0176 & 0.05 & 215 & 0.2 & E\\
			i & 0.2 & 0.8 & 0.8 & 0.2 & 55 & 21 & 23 & 1.56 $\times$ 10$^4$ & 0.0176 & 0.05 & 893 & 0.3 & $\sim$FE\\
			j & 0.8 & 0.2 & 0.8 & 0.2 & 53 & 26 & 28 & 1.56 $\times$ 10$^4$ & 0.0176 & 0.11 & 889 & 0.3 & FE\\
			k & 0.2 & 0.8 & 0.2 & 0.8 & 59 & 30 & 34 & 1.56 $\times$ 10$^4$ & 0.0176 & 0.25 & 800 & 0.2 & FE\\
			l & 1.0 & 0.0 & 0.0 & 1.0 & 44 & 49 & 39 & 1.56 $\times$ 10$^4$ & 0.0176 & 1.00 & 989 & 0.1 & U\\
			m & 1.0 & 0.0 & 0.0 & 1.0 & 51 & 55 & 29 & 1.39 $\times$ 10$^4$ & 0.0159 & 1.00 & 870 & 0.5 & U\\
			n & 1.0 & 0.0 & 0.0 & 1.0 & 57 & 43 & 34 & 1.56 $\times$ 10$^4$ & 0.0176 & 0.25 & 596 & 0.04 & U\\
			o & 1.0 & 0.0 & 0.0 & 1.0 & 39 & 62 & 34 & 1.74 $\times$ 10$^4$ & 0.0193 & 4.00 & 1471 & 0.04 & U\\
		\end{tabular}
	\end{center}
	\caption{Label of tested cases, initial concentration (mass basis) of each species within the target ($\phi_{1t}$ and $\phi_{2t}$) and impact ($\phi_{1i}$ and $\phi_{2i}$) piles, initial diameters of target and impact piles, $D_t$ and $D_i$, respectively, initial separation $\Delta x_d$, channel Reynolds number $Re$, undisturbed shear velocity $u_*$, ratio between initial masses $m_i/m_t$, proposed timescale $t_s$ (Eq. \ref{Eq:timescale2} in Subsection \ref{sec:results_timescale}), characteristic time $t_c$ (shown in Subsection \ref{sec:results_timescale}), and interaction pattern \textit{Pat}. C, M, E, FC, FE and U stand for chasing, merging, exchange, fragmentation-chasing, fragmentation-exchange and undefined patterns, respectively.}
	\label{tab1}
\end{table}

\begin{figure}[h!]
	\begin{center}
		\includegraphics[width=.95\linewidth]{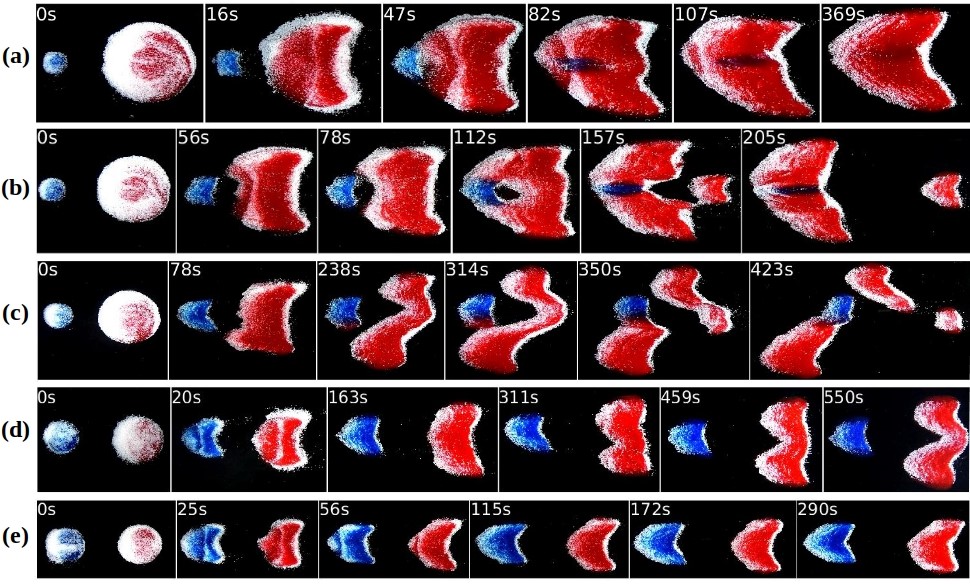}
	\end{center}
	\caption{Snapshots of interactions of bidisperse barchans for fixed initial concentrations $\phi_1$ = 0.5 and $\phi_2$ = 0.5. In the snapshots, the white (clearer) beads correspond to $d_2$ = 0.5 mm and red and blue (darker) beads to $d_1$ = 0.2 mm, the water flow is from left to right, and the corresponding times are shown in each frame. Figures (a) to (e) correspond to cases $a$ to $e$ of Tab. \ref{tab1}: (a) merging; (b) exchange; (c) fragmentation-exchange; (d) fragmentation-chasing; (e) chasing.}
	\label{fig:snap_mixtures}
\end{figure}

\begin{figure}[h!]
	\begin{center}
		\includegraphics[width=.95\linewidth]{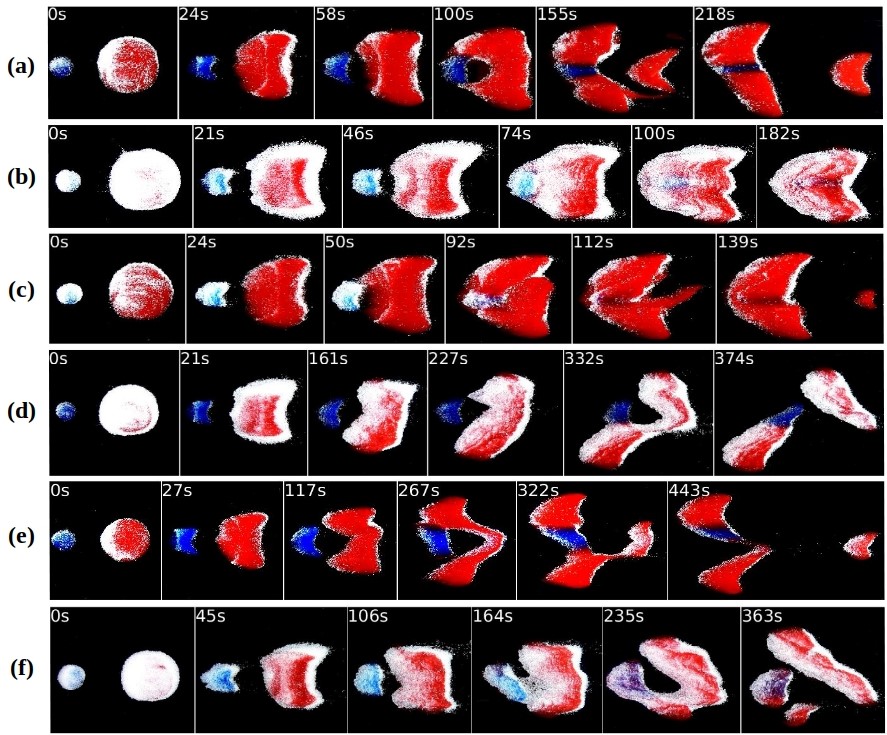}
	\end{center}
	\caption{Snapshots of interactions of bidisperse barchans for fixed initial concentrations $\phi_1$ and $\phi_2$ alternating between either 0.2 or 0.8. In the snapshots, the white (clearer) beads correspond to $d_2$ = 0.5 mm and red and blue (darker) beads to $d_1$ = 0.2 mm, the water flow is from left to right, and the corresponding times are shown in each frame. Figures (a) to (f) correspond to cases $f$ to $k$ of Tab. \ref{tab1}: (a) exchange; (b) merging; (c) exchange; (d) fragmentation-exchange; (e) fragmentation-exchange; (f) fragmentation-exchange.}
	\label{fig:snap_mixtures2}
\end{figure}

We followed the barchans consisting of bidisperse mixtures, for different concentrations of species 1 and 2, and we found patterns similar to those found by Assis and Franklin \cite{Assis} for monodisperse (one-species) barchans. The obtained patterns are shown in Fig. \ref{fig:snap_mixtures} for fixed concentrations $\phi_1$ = 0.5 and $\phi_2$ = 0.5 in both initial piles (cases $a$ to $e$ in Tab. \ref{tab1}), and in Fig. \ref{fig:snap_mixtures2} for concentrations $\phi_1$ and $\phi_2$ alternating between either 0.2 or 0.8 (cases $f$ to $k$ in Tab. \ref{tab1}). Movies for all the cases are available on an open repository \cite{Supplemental2}, and for some cases in the supplementary material.

For $\phi_1$ = $\phi_2$ = 0.5, the same interaction patterns of one-species barchans occur \cite{Assis}, namely the chasing, merging, exchange, fragmentation-chasing and fragmentation-exchange patterns (Figs. \ref{fig:snap_mixtures}(e), \ref{fig:snap_mixtures}(a), \ref{fig:snap_mixtures}(b), \ref{fig:snap_mixtures}(d) and \ref{fig:snap_mixtures}(c), respectively). By alternating $\phi_1$ and $\phi_2$ between either 0.2 or 0.8, using the same (Figs. \ref{fig:snap_mixtures2}(a), \ref{fig:snap_mixtures2}(b), \ref{fig:snap_mixtures2}(e) and \ref{fig:snap_mixtures2}(f), corresponding to cases $f$, $g$, $j$ and $k$ of Tab. \ref{tab1}) or inverted (Figs. \ref{fig:snap_mixtures2}(c) and \ref{fig:snap_mixtures2}(d), corresponding to cases $h$ and $i$ of Tab. \ref{tab1}) concentrations for the impact and target barchans, and for three values of $m_i/m_t$ while all the other parameters were kept constant, we obtained the merging, exchange and fragmentation-exchange patterns. We note that it is probable that the chasing and fragmentation-chasing patterns exist also for the concentrations employed. We did not, however, varied the mass ratio in order to seek for them.

Although the bidisperse piles (mixtures) produce the same interaction patterns observed for one-species barchans, they present some peculiarities in terms of morphodynamics. Two of them are related with grain segregation, as shown by Alvarez et al. \cite{Alvarez5} for single barchans: the accumulation of the smaller grains over the surface of bedforms, and the appearance of a transient stripe, transverse to the flow direction, that initiates upstream the crest of the initial bedform and migrates toward its leading edge until disappearing. Alvarez et al. \cite{Alvarez5} showed that the transient stripe separates the region where segregation is complete from that where segregation is ongoing, and the same feature applies here since bidisperse conical piles are being deformed into bidisperse barchans. Another difference is the formation of a large void (absence of grains) when the impact barchan reaches the target one in the exchange pattern (Figs. \ref{fig:snap_mixtures}(b) and \ref{fig:snap_mixtures2}(a), corresponding to cases $b$ and $f$ in Tab. \ref{tab1}). This void region occurs in the recirculation bubble of the impact barchan and persists until a baby barchan containing only grains from the target one is ejected. The baby barchan has roughly the same projected area of the impact barchan when $\phi_1$ = $\phi_2$ = 0.5, but not when $\phi_1$ $\neq$ $\phi_2$, and, just after the baby barchan is ejected (at 157 s in Fig. \ref{fig:snap_mixtures}(b) and 155 s in Fig. \ref{fig:snap_mixtures2}(a)), the parent bedform has an unusual shape, resembling two elongated barchans containing grains from the target barchan and linked by grains from the impact one. After some time, the parent bedform attains a barchan shape.

\begin{figure}[h!]
	\begin{center}
		\begin{minipage}{0.49\linewidth}
			\begin{tabular}{c}
				\includegraphics[width=0.99\linewidth]{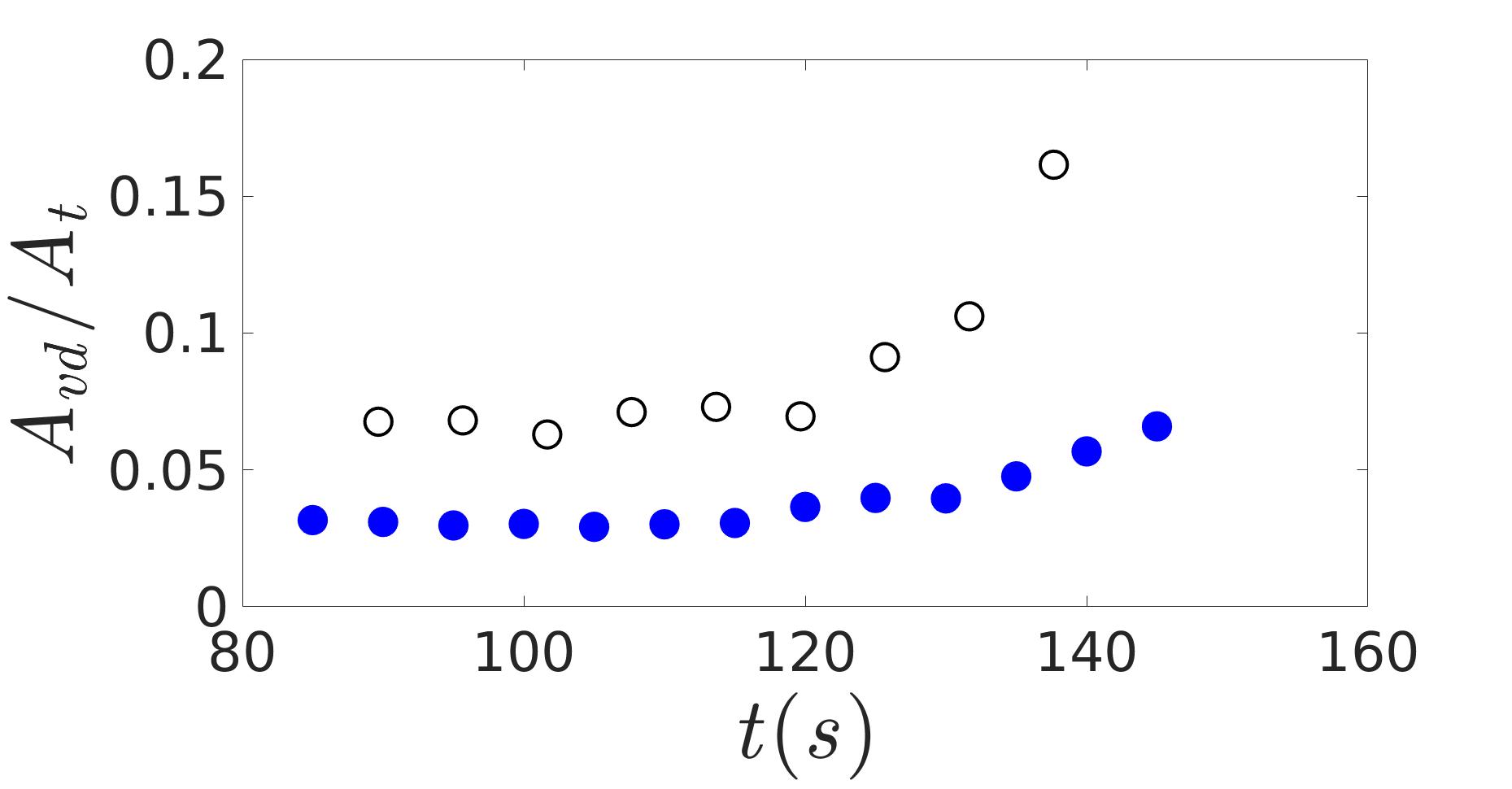}\\
				(a)
			\end{tabular}
		\end{minipage}
		\hfill
		\begin{minipage}{0.49\linewidth}
			\begin{tabular}{c}
				\includegraphics[width=0.99\linewidth]{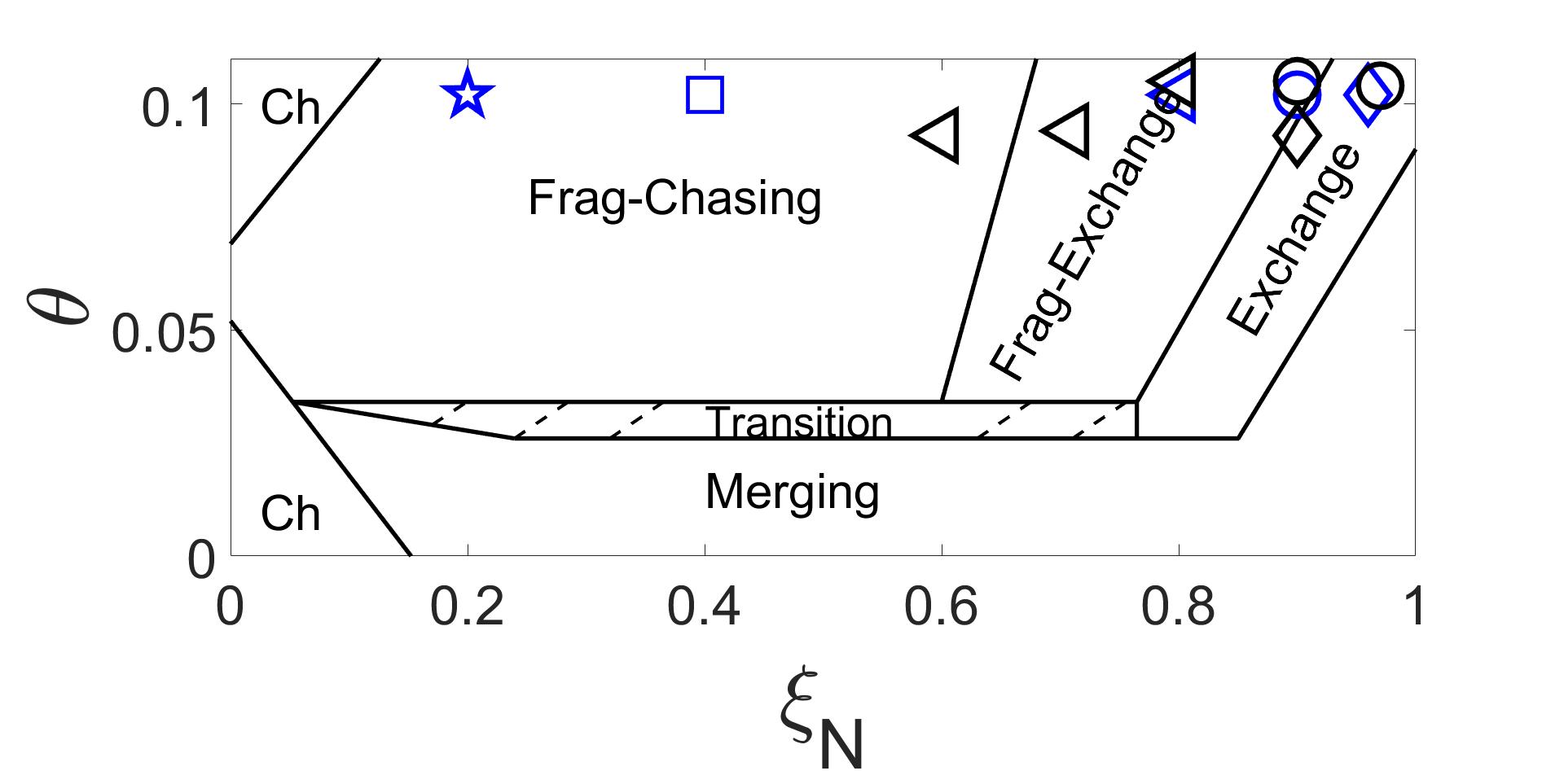}\\
				(b)
			\end{tabular}
		\end{minipage}
		\hfill
	\end{center}
	\caption{(a) Area occupied by the void region $A_{vd}$ normalized by the total area (projected) $A_t$ occupied by grains as a function of time. Solid blue circles correspond to the exchange pattern when $\phi_1$ = $\phi_2$ = 0.5 (Fig. \ref{fig:snap_mixtures}(b), case $b$) and open black symbols when $\phi_1$ $\neq$ $\phi_2$ (Fig. \ref{fig:snap_mixtures2}(a), case $f$). (b) Interaction patterns for barchans of bidisperse mixtures: classification map proposed by Assis and Franklin \cite{Assis} for monodisperse barchans, over which we superposed the experimentally obtained chasing - Ch ($\star$), merging ($\diamond$), exchange ($\circ$), fragmentation-chasing ($\square$), and fragmentation-exchange ($\triangleleft$) patterns for the bidisperse case. Blue color corresponds to $\phi_1$ = $\phi_2$ = 0.5 (Fig. \ref{fig:snap_mixtures}) and black to $\phi_1$ $\neq$ $\phi_2$ (Fig. \ref{fig:snap_mixtures2}). Figure modified from Assis and Franklin \cite{Assis}.}
	\label{fig:mixtures_areas}
\end{figure}

Figure \ref{fig:mixtures_areas}(a) presents the time evolution of areas occupied by the void region $A_{vd}$, normalized by the total area (projected area) $A_t$ occupied by grains, for the exchange cases when $\phi_1$ = $\phi_2$ = 0.5 (Fig. \ref{fig:snap_mixtures}(b), case $b$) and $\phi_1$ $\neq$ $\phi_2$ (Fig. \ref{fig:snap_mixtures2}(a), case $f$). The computation of  areas began when the impact barchan reached the target one, forming a closed void, and finished when the void was no longer closed, the baby barchan being ejected just afterward (see the supplementary material for examples of void region detection and the time evolution of $A_{vd}$ in dimensional form). We observe that the void area corresponds to 5 to 10\% of the projected area occupied by the grains, the void remaining roughly constant for a certain time and then increasing considerably by the time the baby barchan is to be ejected. We observe also that the void is greater when in the presence of a large concentration of smaller grains. The reasons and mechanisms by which the void is formed remain to be investigated further, but they seem associated with granulometric distributions, since we had not observed voids in our experiments with one-species barchans \cite{Assis, Assis2}.

As a consequence of the differences aforementioned, the resulting patterns in the bidisperse case do not necessarily occur under the same conditions as for monodisperse barchans. Figure \ref{fig:mixtures_areas}(b) plots the experimental points measured with bidisperse barchans in the map proposed by Assis and Franklin \cite{Assis} for the aligned case (figure modified from Ref. \cite{Assis}), which is drawn in the parameter space consisting of the Shields number $\theta$ and dimensionless particle number $\xi_N$ = $\Delta_N / \Sigma_N$, where $\Delta_N$ is the difference and $\Sigma_N$ the sum of the number of grains forming each pile. For the bidisperse case (symbols in Fig. \ref{fig:mixtures_areas}(b)), $\theta$ of each pile was computed for each species (i.e., $\theta_1$ and $\theta_2$ using $d_1$ and $d_2$, respectively) and then averaged by the number of grains of each species: $\theta$ =  $N_1\theta_1/N$  $+$ $N_2\theta_2/N$, where $N1$, $N_2$ and $N$ are the number of grains of species 1, 2, and their sum, respectively. In addition, for cases where impact and target barchans had different compositions (cases $h$ and $i$), $\theta$ was afterward computed as an averaged weighted by the size (total number of grains) of each barchan. We observe that most of points fall within the corresponding patterns found in the monodisperse case, or very near the boundaries, but some of them deviate, crossing regions in the map. In general, while the exchange, fragmentation-chasing and fragmentation-exchange tend to remain within their respective boundaries, the chasing and merging patterns deviate considerably, the former crossing the line and occupying part of the fragmentation-chasing region and the latter occupying part of the exchange region. We note that if we had used $m_i/m_t$ or a length ratio (as other authors did \cite{Endo2, Katsuki2, Katsuki, Genois}) instead of $\xi_N$, many of the symbols would be superposed in Fig. \ref{fig:mixtures_areas}(b) even if measured patterns are different, since the mass ratio does not take into consideration details about granular compositions. This corroborates, in a certain way, the use of a dimensionless parameter based on the number of elements ($\xi_N$) rather than $m_i/m_t$.

\begin{figure}[h!]
	\begin{center}
		\includegraphics[width=.5\linewidth]{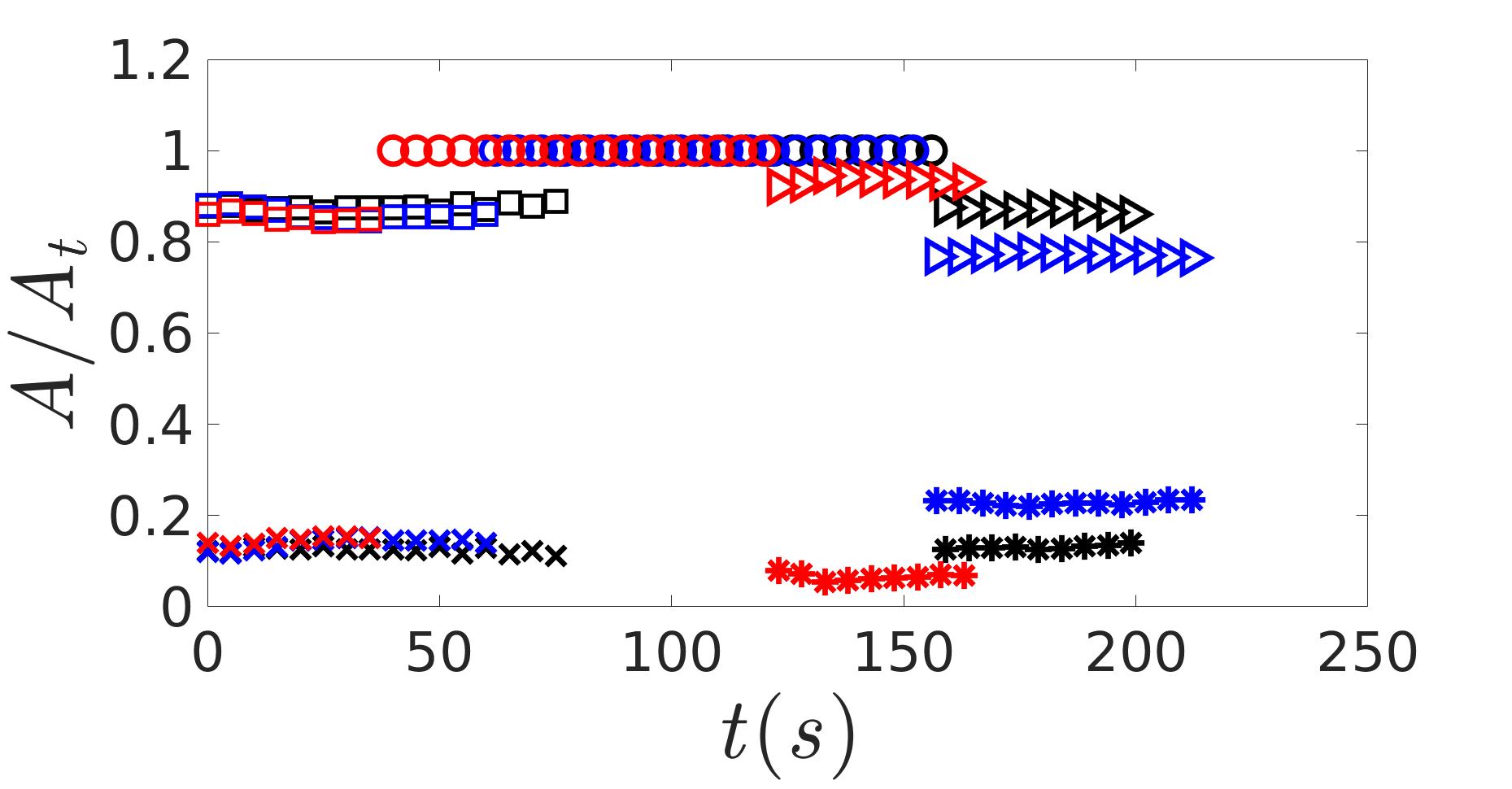}
	\end{center}
	\caption{Projected areas, normalized by $A_t$, of impact (x), target ($\Box$), merged ($\circ$),  parent ($\triangleright$), and baby (*) barchans, respectively, as functions of time, during exchange processes. Black, blue and red colors correspond to cases $a$, $f$ and $h$ (Figs. \ref{fig:snap_mixtures}b, \ref{fig:snap_mixtures2}a and \ref{fig:snap_mixtures2}c), respectively.}
	\label{fig:areas_exchance}
\end{figure}

In particular, we analyzed the projected areas occupied by grains in the case of exchange patterns. For monodisperse dunes, Assis and Franklin \cite{Assis, Assis2} showed that the impact barchan first merges with the target one, and afterward a new barchan is ejected, known as baby barchan, the remaining bedform being the parent barchan. In addition, Assis and Franklin \cite{Assis} showed that the baby barchan has roughly the same size of the impact barchan, but contains grains only from the target one. In the case of bidisperse mixtures, the same behavior happens. Figure \ref{fig:areas_exchance} shows the projected areas of bedforms during the exchange processes of Figs. \ref{fig:snap_mixtures}(b), \ref{fig:snap_mixtures2}(a) and \ref{fig:snap_mixtures2}(c) (cases $a$, $f$ and $h$, corresponding to black, blue and red colors, respectively. A graphic in dimensional form is available in the supplementary material). While the baby barchan does not contain grains from the impact barchan (seen directly from Figs. \ref{fig:snap_mixtures}(b), \ref{fig:snap_mixtures2}(a) and \ref{fig:snap_mixtures2}(c)), Fig. \ref{fig:areas_exchance} shows that their areas are roughly the same.

In summary, we show that in the mixed case the interaction patterns and their dynamics are roughly the same as in the monodisperse case; however, although the map proposed in Ref. \cite{Assis} brings valuable information for classifying the barchan-barchan interactions, results with polydisperse dunes can deviate from the proposed boundaries. Therefore, the distribution of grains within the barchans should be taken into consideration in analyses of barchan-barchan interactions occurring in nature.

\subsection{Two-species monodisperse piles}
\label{sec:results_monodisp}

\begin{figure}[h!]
	\begin{center}
		\includegraphics[width=.95\linewidth]{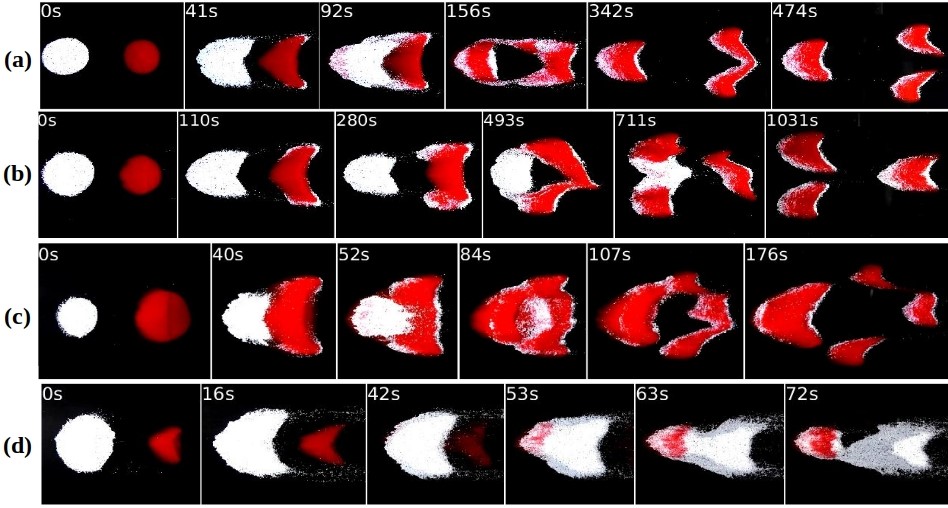}
	\end{center}
	\caption{Snapshots of barchan interactions for initially monodisperse piles of different grains (two-species monodisperse piles). In the snapshots, the upstream pile consists of white (clearer) beads with $d_2$ = 0.5 mm and the downstream pile of red (darker) beads with $d_1$ = 0.2 mm, the water flow is from left to right, and the corresponding times are shown in each frame. Figures (a) to (d) correspond to cases $l$ to $o$ of Tab. \ref{tab1}.}
	\label{fig:snap_mono}
\end{figure}

We followed the initially monodisperse bedforms consisting each of different grains, and we found different patterns. These patterns, cases $l$ to $o$ in Tab. \ref{tab1}, are shown in Fig. \ref{fig:snap_mono}, which presents snapshots of barchans at some instants during their interactions, including a collision where the impact barchan was larger than the target one (see the supplementary material or Ref. \cite{Supplemental2} for movies of collisions). For this specific case, we made use of larger grains in the impact dune since the displacement velocity of barchans varies with the diameter of their grains \cite{Franklin_8} (see Eq. \ref{Eq:barchan_velocity} in Subsection \ref{sec:results_timescale}).

\begin{figure}[h!]
	\begin{center}
		\begin{minipage}{0.49\linewidth}
			\begin{tabular}{c}
				\includegraphics[width=0.99\linewidth]{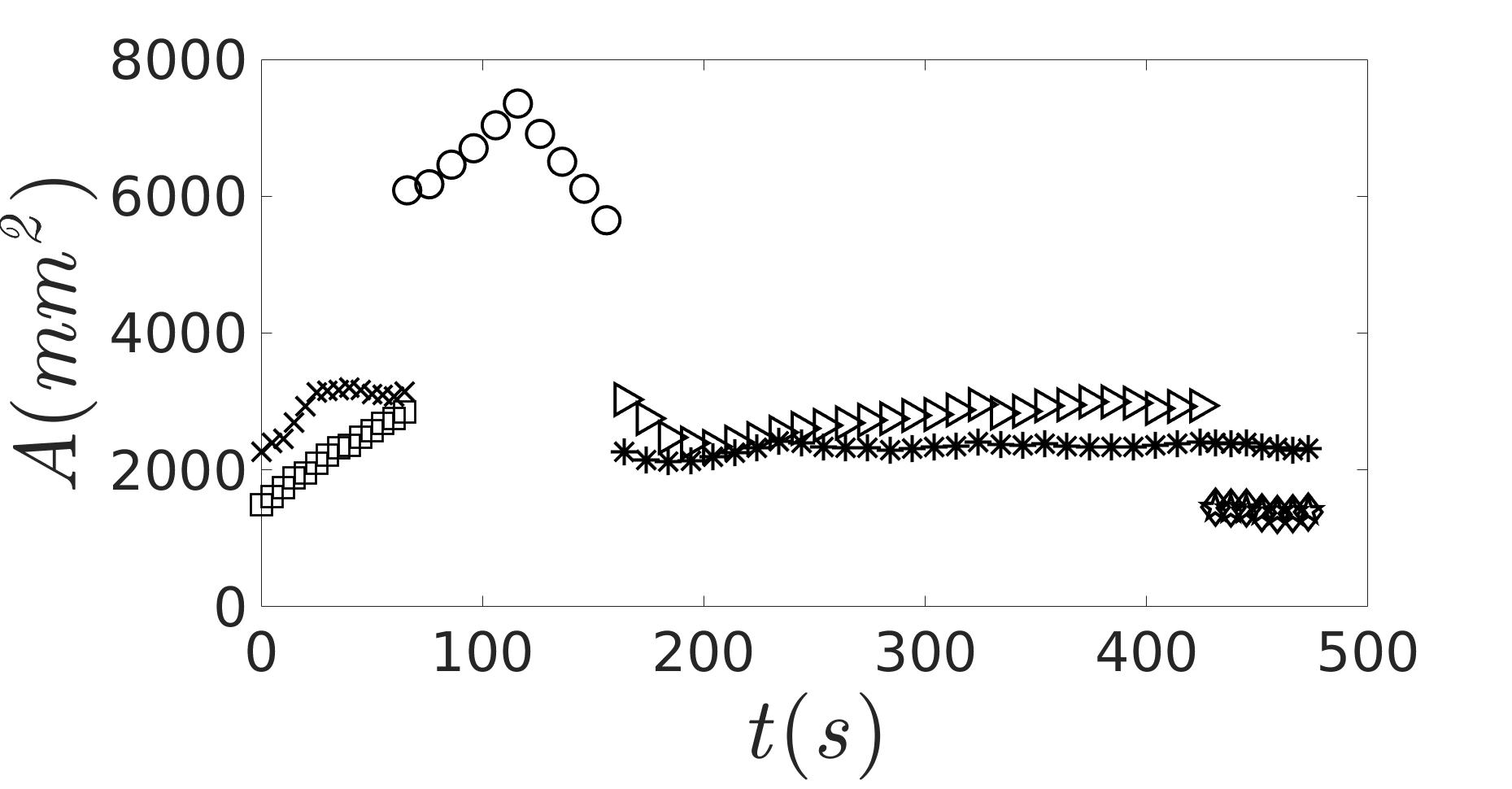}\\
				(a)
			\end{tabular}
		\end{minipage}
		\hfill
		\begin{minipage}{0.49\linewidth}
			\begin{tabular}{c}
				\includegraphics[width=0.99\linewidth]{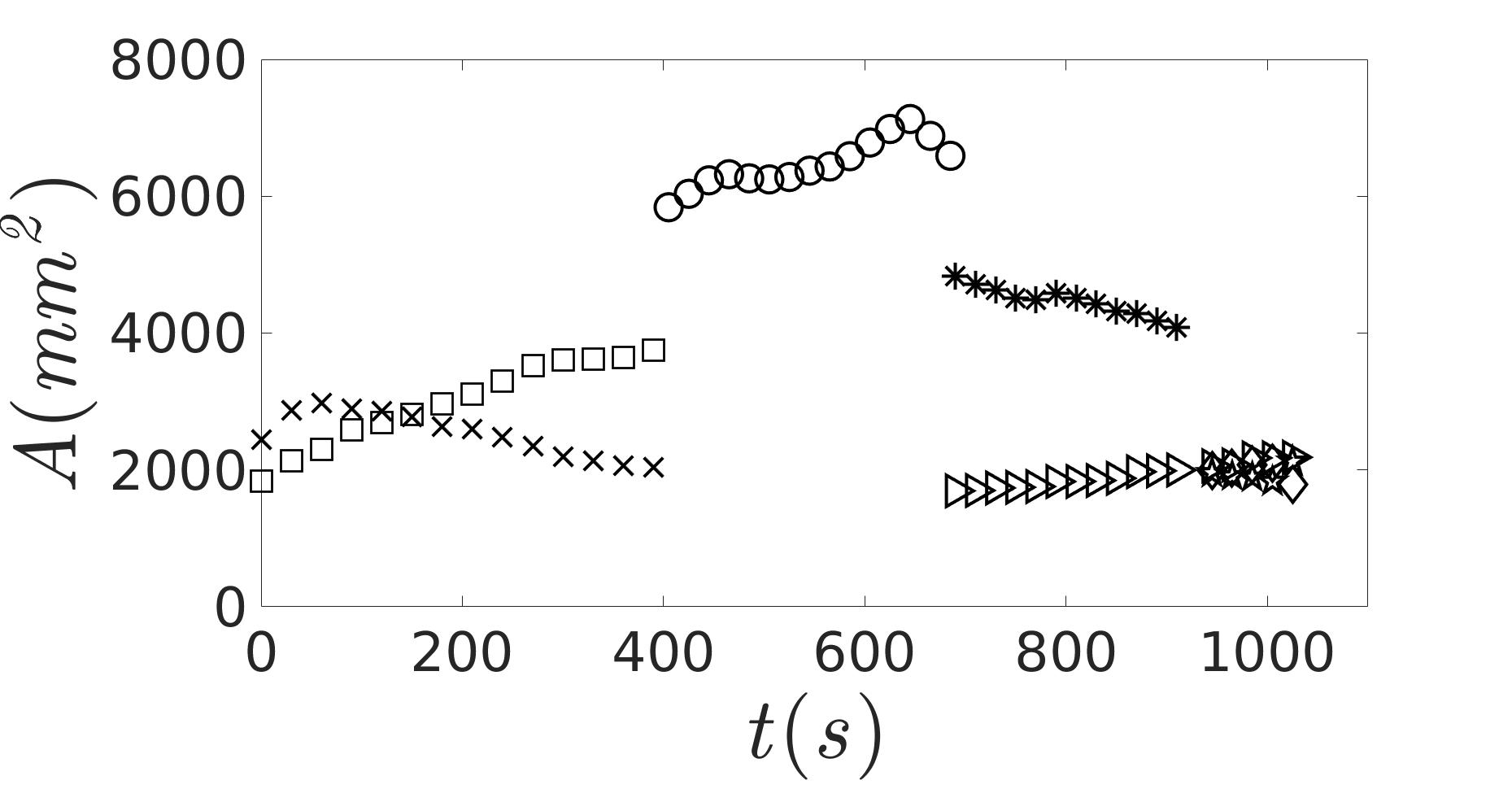}\\
				(b)
			\end{tabular}
		\end{minipage}
		\hfill
				\begin{minipage}{0.49\linewidth}
			\begin{tabular}{c}
				\includegraphics[width=0.99\linewidth]{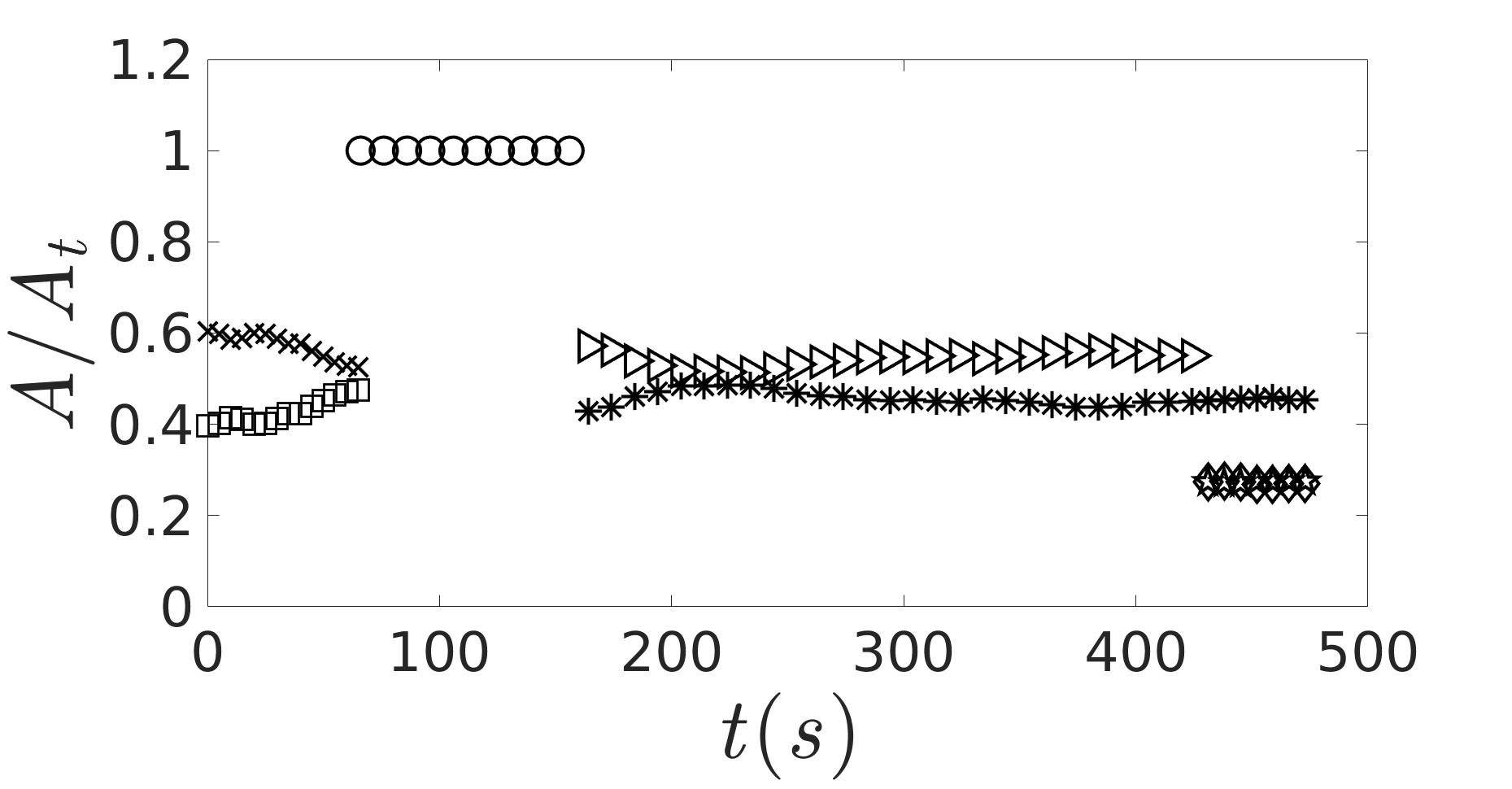}\\
				(c)
			\end{tabular}
		\end{minipage}
		\hfill
		\begin{minipage}{0.49\linewidth}
			\begin{tabular}{c}
				\includegraphics[width=0.99\linewidth]{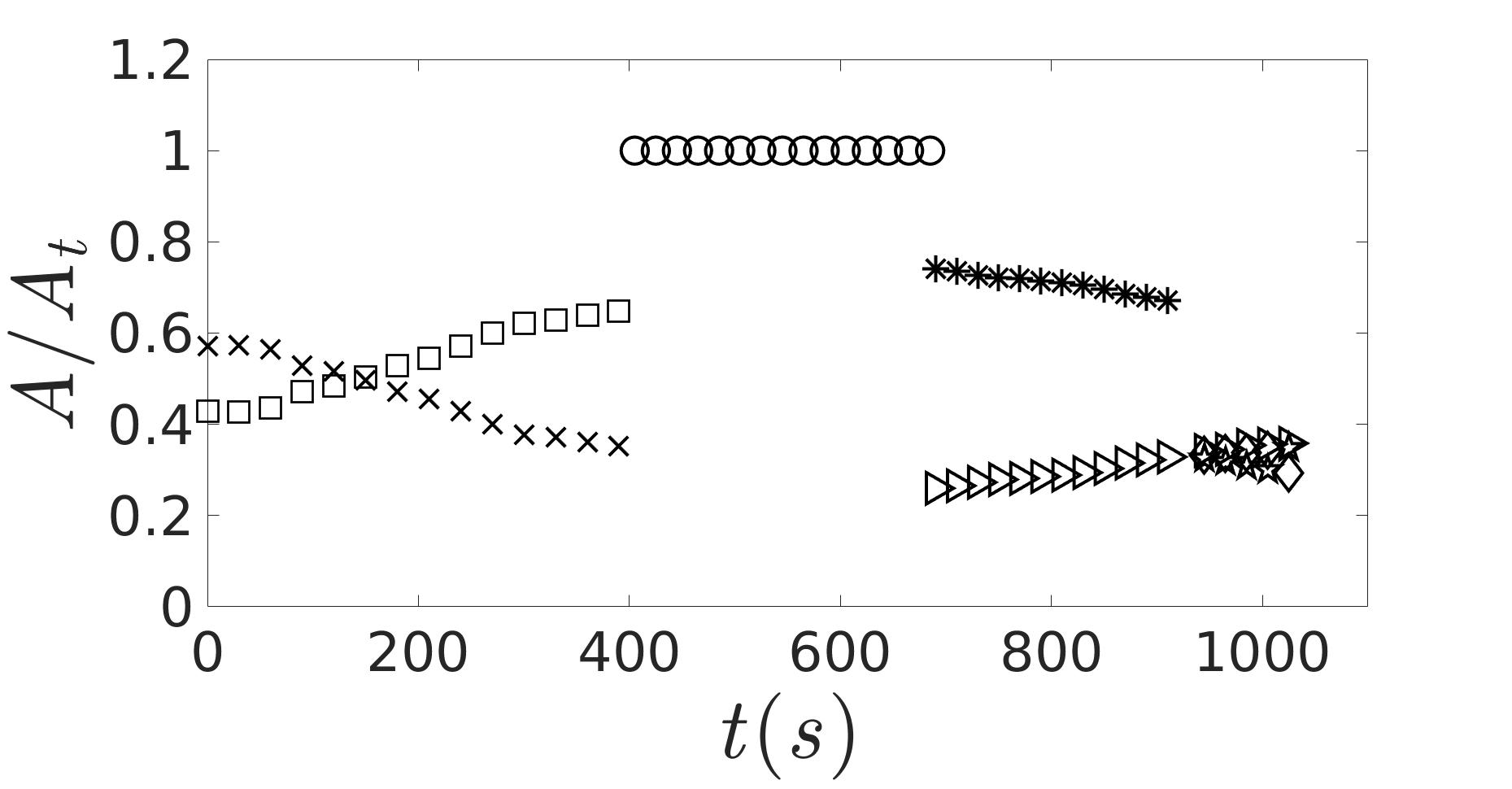}\\
				(d)
			\end{tabular}
		\end{minipage}
		\hfill
	\end{center}
	\caption{Projected areas of impact (x), target ($\Box$), merged ($\circ$), upstream after splitting (*), parent ($\triangleright$), and baby barchans ($\diamond$ and $\star$), respectively, during undefined-exchange processes for (a) and (c) case $l$ (Fig. \ref{fig:snap_mono}(a)) and (b) and (d) case $m$ (Fig. \ref{fig:snap_mono}(b)). Figures (a) and (b) are in dimensional and (c) and (d) in dimensionless form (normalized by $A_t$).}
	\label{fig:mono_areas}
\end{figure}

For all the interactions shown in Fig. \ref{fig:snap_mono}, the impact barchan consisted of grains of species 2 ($d_2$ = 0.5 mm) in white color and the target barchan of species 1 ($d_1$ = 0.2 mm) in red color. In Fig. \ref{fig:snap_mono}(a) (case $l$ in Tab. \ref{tab1}), the barchans collide, forming a large void in the recirculation region when they touch each other, and with the larger grains moving over the smaller ones that, in their turn, emerge at the toe of the resulting bedform (at 92 s). Afterward (at 156 s), the smaller grains having accumulated over the surface of the resulting bedform, the latter begins to split in what seems, initially, two barchans linked by two large branches. Finally (at 474s), they split in three barchans (one upstream and two downstream, in a staggered configuration) consisting each of bidisperse grains. Figures \ref{fig:mono_areas}(a) and \ref{fig:mono_areas}(b) show the time evolution of projected areas of dunes for cases $l$ and $m$, respectively (Figs. \ref{fig:mono_areas}(c) and \ref{fig:mono_areas}(d) in dimensionless form, normalized by $A_t$). Interestingly, we can observe from Fig. \ref{fig:mono_areas}(a) that, after the dunes collide (at $\sim$ 90 s), the projected area of the resulting bedform first increases and then decreases. This is due, respectively, to larger grains migrating over the smaller ones and spreading over the dune, and afterward the smaller grains accumulating over the dune surface and decreasing the projected area. In addition, Fig. \ref{fig:mono_areas}(a) shows that the two final downstream barchans have roughly the same size.

In Fig. \ref{fig:snap_mono}(b) (case $m$ in Tab. \ref{tab1}), the behavior is similar to that of Fig. \ref{fig:snap_mono}(a), the main difference being that the final state is inversed: two upstream barchans and one downstream barchan, in a staggered configuration. In both cases $l$ and $m$ the initial masses are the same (the initial diameter of the impact pile being larger since it consists of larger grains), only the fluid velocity is different, being higher for case $l$. Why the behavior changes by changing the water velocity remains to be investigated (we do not advance an explanation for the moment). However, we can observe from Fig. \ref{fig:mono_areas}(b) the same increase and decrease of the projected area after the dunes have collided (at $\sim$ 400 s), that the resultant bedform splits first in one larger upstream and one smaller downstream bedform (at $\sim$ 700 s), and that the upstream bedform splits in two dunes later (at $\sim$ 950 s). We end finally with three barchans of roughly the same size.

Figure \ref{fig:snap_mono}(c) corresponds to case $n$ of Tab. \ref{tab1}, and shows a collision in which three barchans are ejected from the merged bedform. This case resembles the exchange pattern \cite{Assis}, with the difference that three baby barchans \cite{Assis, Assis2} are ejected, one aligned and two in staggered configuration.

Finally, Fig. \ref{fig:snap_mono}(d) corresponds to case $o$ of Tab. \ref{tab1}, the unusual case of a collision of a larger impact with a smaller target barchan (as far as we know, this is the first time that this kind of collision is reported. See the supplementary material or Ref. \cite{Supplemental2} for a movie of this interaction). We observe that, as the impact barchan gets closer to the target one, the latter becomes more elongated, while grains leaving the horns of the impact barchan are entrained further downstream and are not incorporated by the target barchan. During the collision (t $\approx$ 48 s), the larger grains move over the smaller ones, which, in their turn, emerge at the toe of the resulting bedform. Finally, a monodisperse baby barchan consisting of only larger grains is ejected from the merged bedform (grains from the impact dune, different from all cases reported previously), resulting, in fact, of larger grains being entrained further downstream. The remaining grains form an upstream bidisperse barchan with smaller grains populating its upper surface.

\begin{figure}[h!]
	\begin{center}
		\begin{minipage}{0.49\linewidth}
			\begin{tabular}{c}
				\includegraphics[width=0.99\linewidth]{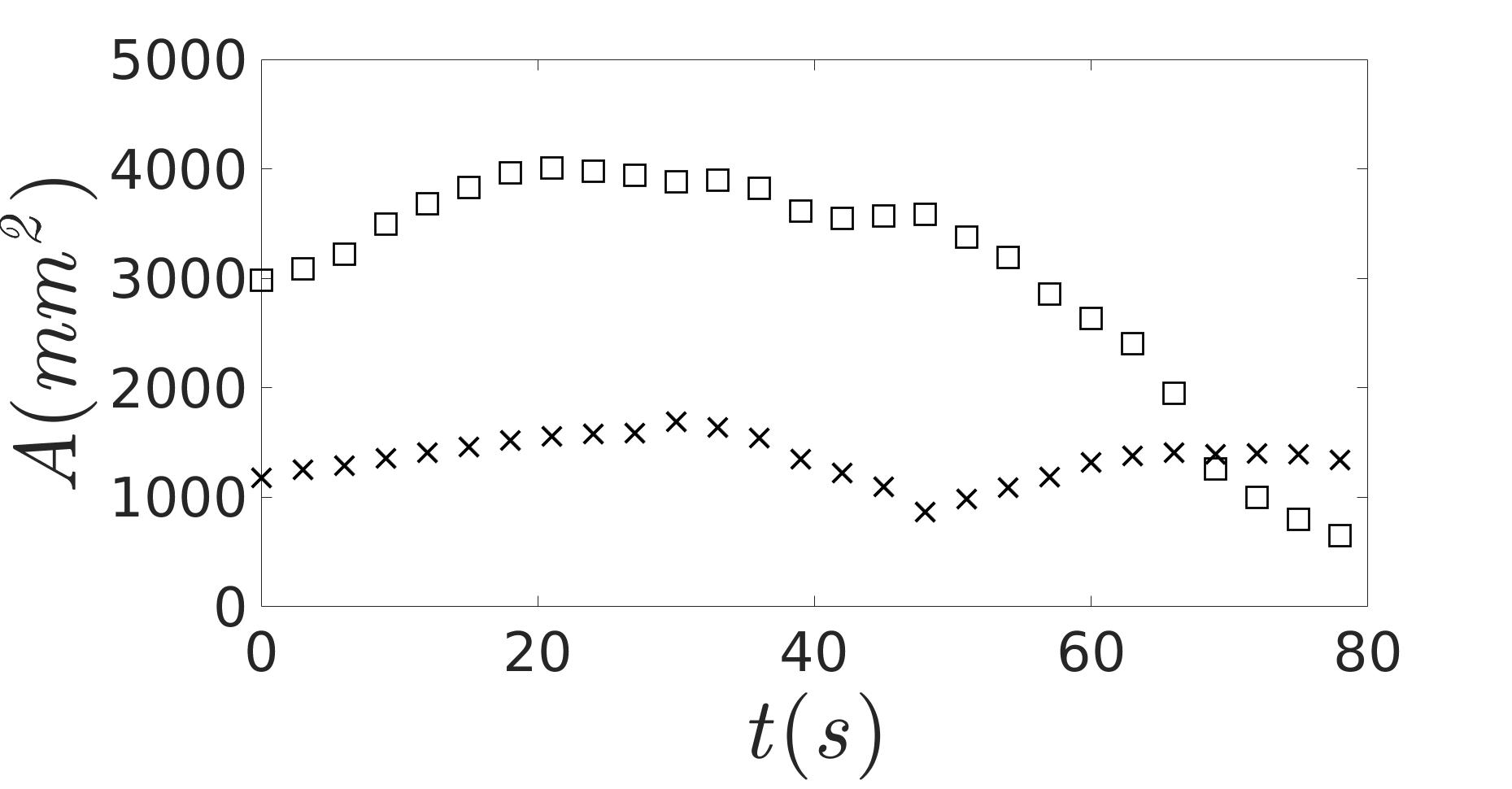}\\
				(a)
			\end{tabular}
		\end{minipage}
		\hfill
		\begin{minipage}{0.49\linewidth}
			\begin{tabular}{c}
				\includegraphics[width=0.99\linewidth]{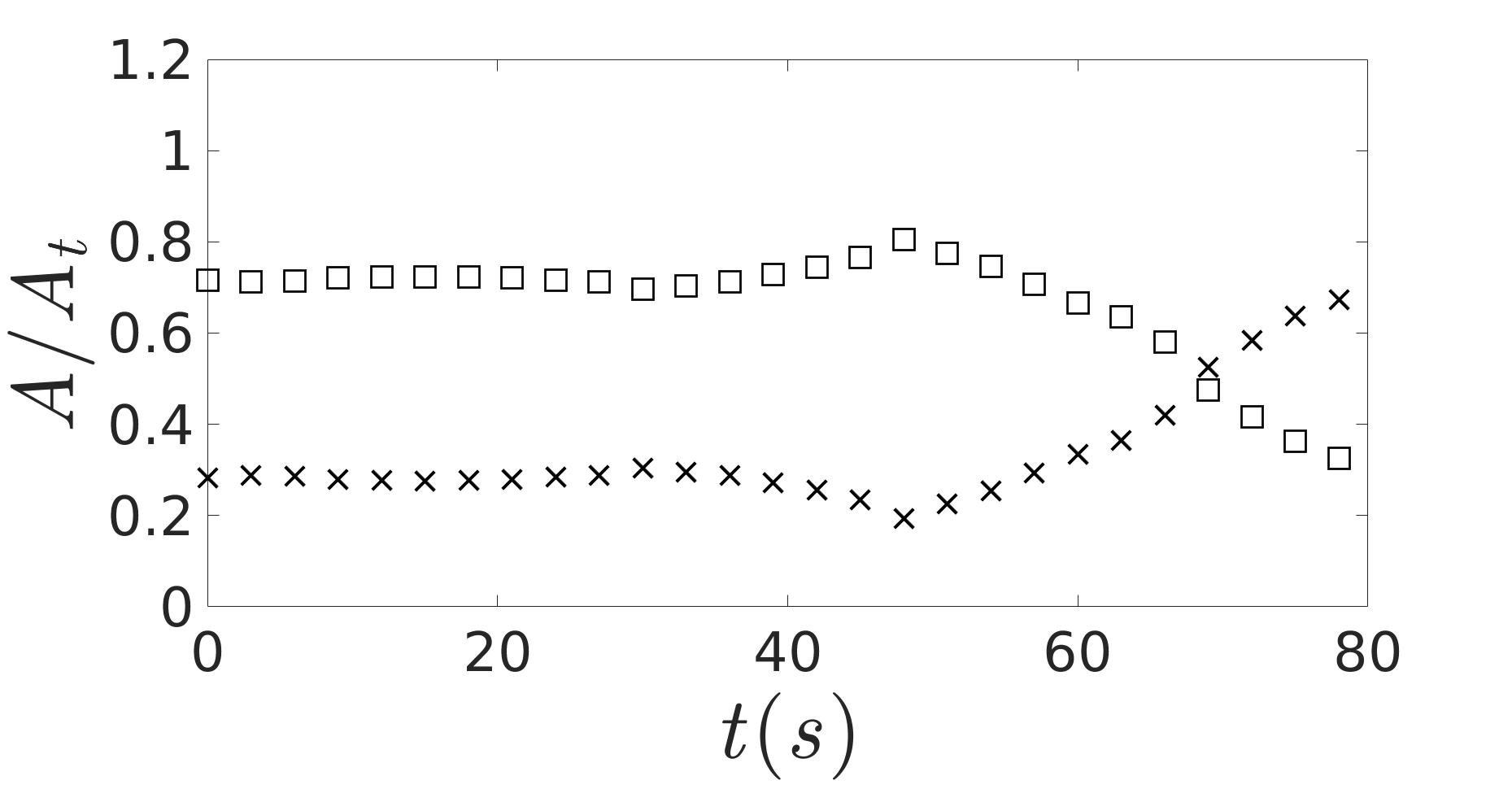}\\
				(b)
			\end{tabular}
		\end{minipage}
		\hfill
	\end{center}
	\caption{Time evolution of projected areas occupied by white ($\square$) and red (x) grains for case $o$ (Fig. \ref{fig:snap_mono}(d)), in (a) dimensional form and (b) normalized by $A_t$. OBS: the monolayer consisting of white grains observed in Fig. \ref{fig:snap_mono}(d) was neglected.}
	\label{fig:area_larger_impact}
\end{figure}

Figure \ref{fig:area_larger_impact} shows the time evolution of projected areas occupied by white and red grains for case $o$. From Fig. \ref{fig:area_larger_impact}(a), we observe an initial increase of both areas due to the spreading of the initial conical pile (being deformed into barchan dunes), with a time interval when areas remain roughly constant. When the larger dune approaches the smaller one, the vortex on the recirculation region of the former entrains grains from the latter toward its lee face. Because the larger grains move easier over the smaller ones \cite{Alvarez5}, these remain on the bottom of the impact barchan until they emerge at the toe of the impact one, being again exposed to the fluid flow. When the dunes collide, the white grains then move over the red (smaller) ones, which appear at the toe of the resulting bedform. This ``swallowing'' process appears in Fig. \ref{fig:area_larger_impact} as a decrease followed by an increase of the area occupied by red grains within 30 s $<$ $t$ $<$ 70 s. Finally, by the end of the collisional process ($t$ $\approx$ 70 s on), a greater number of larger (white) grains are exposed to the fluid flow and are either entrained further downstream or form a monolayer between both barchans, which makes the area occupied by the larger (white) grains to decrease considerably (the monolayer being neglected in Fig. \ref{fig:area_larger_impact}).

In common for cases $l$ to $o$, the larger grains move over the smaller ones, being entrained further downstream and/or accumulating on the lee face and forming a carpet for the smaller grains. The same mechanism proposed by Alvarez et al. \cite{Alvarez5} seems to apply here, i.e., the segregation results from a competition between fluid entrainment and easiness of rolling. Other than that, we do not advance more explanations for the patterns shown in Fig. \ref{fig:snap_mono}, but we propose, however, a timescale that applies to all cases investigated and is presented in Subsection \ref{sec:results_timescale}.

\subsection{Timescale}
\label{sec:results_timescale}

A question that remained to be answered for barchan-barchan interactions even in monodisperse conditions \cite{Endo2, Hersen_5, Assis, Assis2}, and that can now be investigated for bidisperse barchans, is the existence of a proper timescale for the problem. A reasonable timescale for the binary interaction of dunes can be built as their initial separation $\Delta x_d$ divided by their relative velocity $\Delta V_d$,

\begin{equation}
	t_{s} = \frac{\Delta x_d}{\Delta V_d}
	\label{Eq:timescale}
\end{equation}

\noindent where $\Delta V_d$ is the difference between the displacement velocities $V_d$ of impact and target barchans. $t_s$ represents a typical time for the collision of barchans (faster for closer barchans with stronger relative velocities). Franklin and Charru \cite{Franklin_8} investigated the displacement velocities of subaqueous barchans by varying water velocities, grain types, and dune sizes, and found that they vary as

\begin{equation}
	\frac{V_d}{V_{ref}} \sim \frac{d}{L} \left( \theta - \theta_{th} \right)^n
	\label{Eq:barchan_velocity}
\end{equation}

\noindent where $d$ is the mean grain diameter, $L$ is the barchan length, $\theta_{th}$ is the threshold value of the Shields number for incipient motion, $n$ is an exponent, and $V_{ref}$ = $((S-1)gd)^{1/2}$, with $S$ = $\rho_s/\rho$. Observing that the diameter of the initial pile $D$ is proportional to $L$, and considering the mean diameter and density as in Alvarez et al. \cite{Alvarez5}, $\bar{d}$ = $\left( \phi_{1}/d_1 + \phi_{2}/d_2 \right) ^{-1}$ and $\bar{\rho}_s$ = $\phi_1 \rho_1 + \phi_2 \rho_2$, respectively, we obtain

\begin{equation}
	\Delta V_d \sim u_* \left| \frac{\bar{d_t}}{D_t} - \frac{\bar{d}_i}{D_i}\right| \frac{1}{S}
	\label{Eq:relative_velocity}
\end{equation}

\noindent where $\bar{d}_i$ and $\bar{d}_t$ represent the mean diameter of grains forming impact and target barchans, respectively, and $D_i$ and $D_t$ are the initial diameters of the projected areas of impact and target barchans, respectively. Finally,

\begin{equation}
	t_{s} = \frac{\Delta x_d S}{u_*} \left| \frac{\bar{d_t}}{D_t} - \frac{\bar{d}_i}{D_i}\right|^{-1}
	\label{Eq:timescale2}
\end{equation}

By normalizing the time interval that barchans take to complete their interaction $\Delta t$ by the proposed timescale $t_s$, we obtain the characteristic time for interactions $t_c$ = $\Delta t/t_s$ (values shown in Tab. \ref{tab1}). The initial instant for $\Delta t$ is when the flow starts and the final instant when the interaction reaches a stage characteristic of the considered pattern. In cases with collision (merging, exchange and fragmentation-exchange), the final instant is when collision takes place, and in cases without collision (chasing and fragmentation-chasing), the final time is much larger than the duration of tests and we consider $\Delta t$ as tending to infinity. In general, we observe the following characteristic times for interactions:



\begin{itemize}
	\item 0.04 $\leq$ $t_c$ $<$ 2 for cases with collisions, i.e., the merging, exchange, fragmentation-exchange (by considering also the one-species monodisperse barchans presented in Refs. \cite{Assis, Assis2}) and the undefined patterns of the two-species monodisperse barchans;
	\item $t_c$ = $\infty$ for the chasing and fragmentation-chasing patterns.
\end{itemize}

Because only cases $n$ and $o$ (two-species monodisperse barchans) present $t_c$ = 0.04 $<$ 0.1 and only one fragmentation-exchange pattern for one-species monodisperse barchans \cite{Assis2} presents $t_c$ = 1.7 $>$ 1, we obtained, in general, that $t_c$ = $O(0.1)$, where $O()$ stands for "order of magnitude". This characteristic time holds for the barchans of same monodisperse composition (one species) presented in Refs. \cite{Assis, Assis2}, for the barchans consisting of bidisperse mixtures, and for the barchans consisting of different monodisperse grains.

If the proposed timescale proofs to be valid for other environments, it will allow the prediction of durations of barchan-barchan interactions and, more generally, provide a scaling for the evolution of dune fields on terrestrial deserts and other planetary environments.

\section{CONCLUSIONS}
\label{sec:conclusions}

In this paper, we investigated experimentally the dune-dune interactions for barchans consisting of (i) bidisperse mixtures and (ii) different monodisperse grains (one type for each barchan). The experiments were conducted in a water channel where two barchans interacted with each other while filmed by a camera, and the bedform morphologies and duration of interactions were obtained from image processing. We observed that a transient stripe appears over the dunes in cases of bidisperse mixtures (as also happens for single bidisperse barchans \cite{Alvarez5}), that interaction patterns vary with concentrations, and that different interactions exist when each barchan consists of different monodisperse grains (two-species monodisperse barchans). For the latter, we obtained, for the first time, one very peculiar case by using larger grains in the impact barchan: the collision of a larger upstream barchan with a smaller downstream one, which showed a different grain distribution in the resulting bedform once the collision had taken place. Finally, we proposed a timescale for the interactions of both monodisperse and bidisperse barchans (cases i and ii). The identification of such timescale represents a new step for predicting the duration of binary interactions and, more generally, scaling the evolution of dune fields on Earth, Mars and other planetary environments.

\section*{SUPPLEMENTARY MATERIAL}
See the supplementary material for  a brief description of the employed methods, the layout of the experimental setup, microscopy images of the employed grains, additional tables and graphics, and movies of barchan-barchan interactions.

\section*{DATA AVAILABILITY}
The data that support the findings of this study are openly available in Mendeley Data at http://dx.doi.org/10.17632/sbjtzbzh9k.

\begin{acknowledgments}
The authors are grateful to FAPESP (Grant Nos. 2016/18189-0, 2018/14981-7 and 2019/10239-7) for the financial support provided.
\end{acknowledgments}

\bibliography{references}

\clearpage

\begin{center}
	\textbf{SUPPLEMENTARY MATERIAL: Revealing the intricate dune-dune interactions of bidisperse barchans}
\end{center}

\noindent\textbf{\large Introduction}

This supplementary material presents a brief description of the employed methods, the layout of the experimental device, microscopy images of the used grains, additional graphics, and movies showing examples of interactions of bidisperse barchans. For the latter, we present top view movies for barchans consisting of (i) bidisperse mixtures of grains (file caseb.gif) and (ii) different monodisperse grains (one type for each barchan, file caseo.gif). We note that complete tables, individual images and movies used in the manuscript are available on Mendeley Data (http://dx.doi.org/10.17632/sbjtzbzh9k).
\\

\noindent\textbf{\large Methods}

\noindent\textbf{Preparation of experiments}

The solid particles used in the experiments were glass spheres (see microscopy images in Figs. \ref{fig2} to \ref{fig4} below) with diameters 0.15 mm $\leq$ $d_{s1}$ $\leq$ 0.25 mm and 0.40 mm $\leq$ $d_{s2}$ $\leq$ $0.60$ mm (from Sigmund Lindner company). Prior to each test, they were separated and weighted with a precision scale with a resolution of 0.01 g in order to assure the right proportions of grains forming each initial pile, as well as the total mass of the initial pile.  Once weighted, the samples were placed manually in the test section of the channel, already filled with water. With the piles placed on the bottom wall of the channel, a controlled flow of water was imposed and the piles deformed into two barchan dunes that interacted with each other. The desired water flow was fixed manually through globe valves, and the volumetric flow rate was measured with an electromagnetic flow meter (KROHNE, model Optiflux 2010C, 0.5 \% uncertainty, maximum measurement capacity of 20 m3/h). A Nikon D7500 camera (which has maximum resolution of 1980 px $\times$ 1080 px at 60 Hz) with a lens of 18-140 mm focal distance and F2.8 maximum aperture was mounted on a traveling system and had a top view of the bedforms. The focus was adjusted manually and the pixel to millimeter conversion was carried out by placing a scale in the channel (filled with water) and acquiring a calibration image. In order to obtain the necessary light while avoiding beating with the camera frequency, two lamps of light-emitting diode (LED) with 100W each were used.

\noindent\textbf{Image processing}

Once the test run was concluded, the corresponding video file was cropped into frames that were saved as single image files by using functions existing in the Matlab software. Because of the timescales involved, we processed images corresponding to every 1 s in all tests. The image processing began by converting RGB images in grayscale, and then, using a threshold adjusted manually, into binary images. In order to remove small objects and noise, Matlab built-in filters were used (namely the medfilt2 and bwareaopen functions). After that, some morphological information of identified objects, such as the area, width, length, and centroid positions, were obtained with the built-in function regionprops. Properties related to the interacting barchans, such as their width, total length, length of horns, instantaneous separation, etc., were computed with scripts written by ourselves. The process just described was performed inside a loop, which reads and stores the data of every processed image in vectors that are saved in mat files. Finally, mat files are post-processed in order to obtain time evolution of areas, lengths, celerities, etc.

\begin{figure}[b!]
	\begin{center}
		\includegraphics[width=0.95\linewidth]{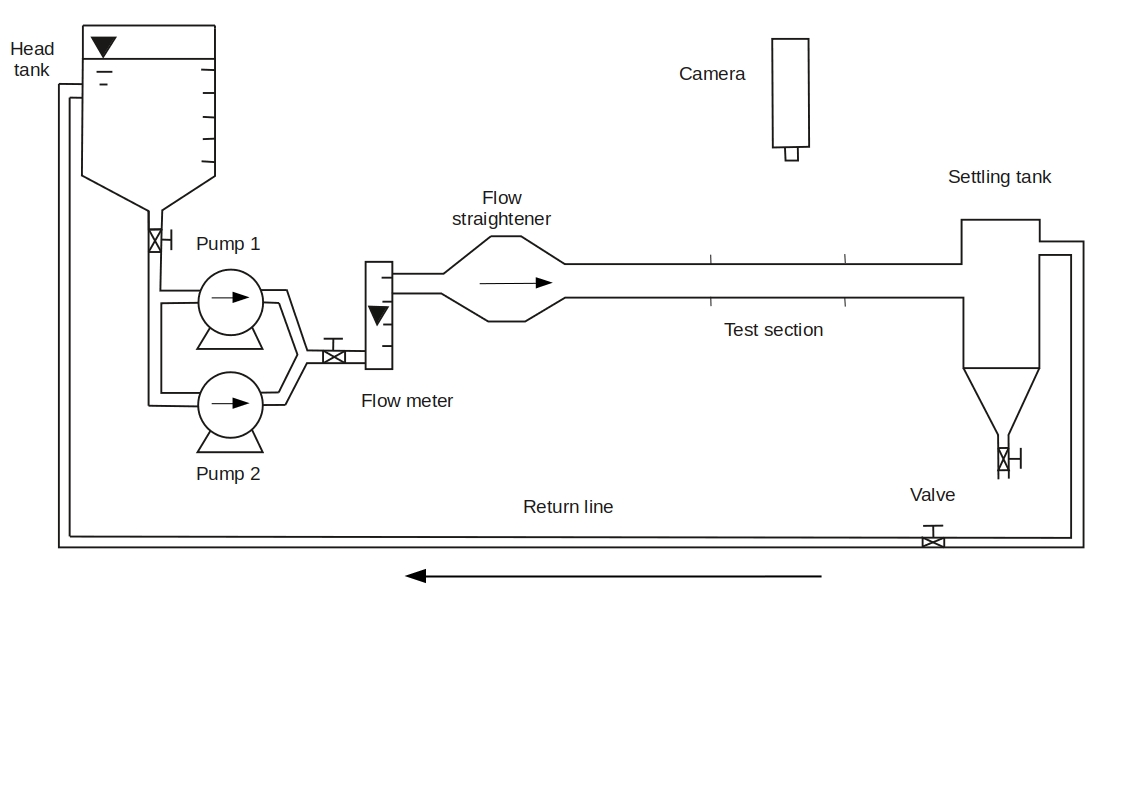}
	\end{center}
	\caption{Layout of the experimental setup.}
\end{figure}

\begin{figure}[b!]
	\begin{center}
		\includegraphics[width=0.7\linewidth]{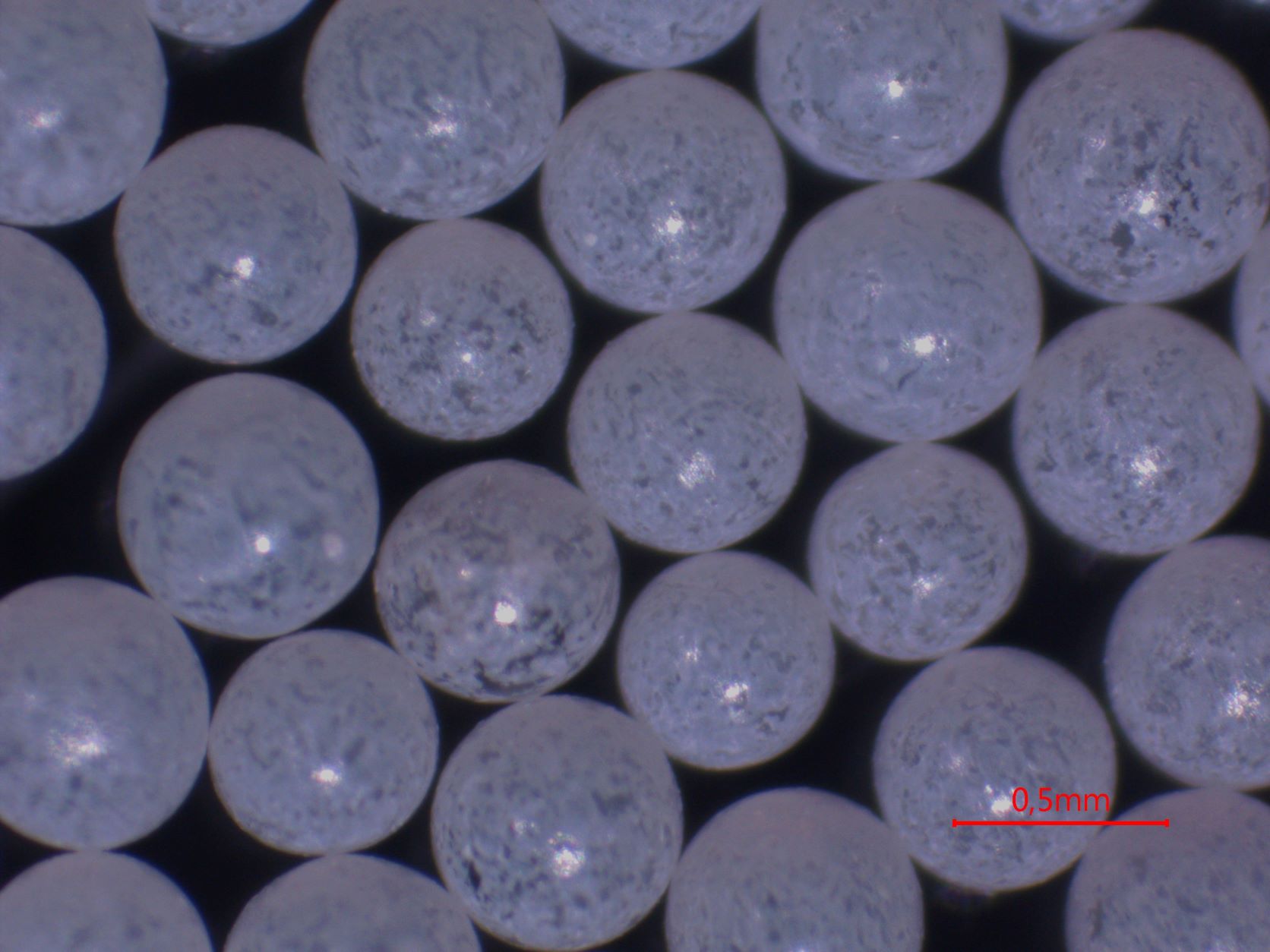}
	\end{center}
	\caption{Microscopy image for the 0.40 mm $\leq$ $d$ $\leq$ 0.60 mm round glass beads of white color (species 2).}
	\label{fig2}
\end{figure}

\begin{figure}[b!]
	\begin{center}
		\includegraphics[width=0.7\linewidth]{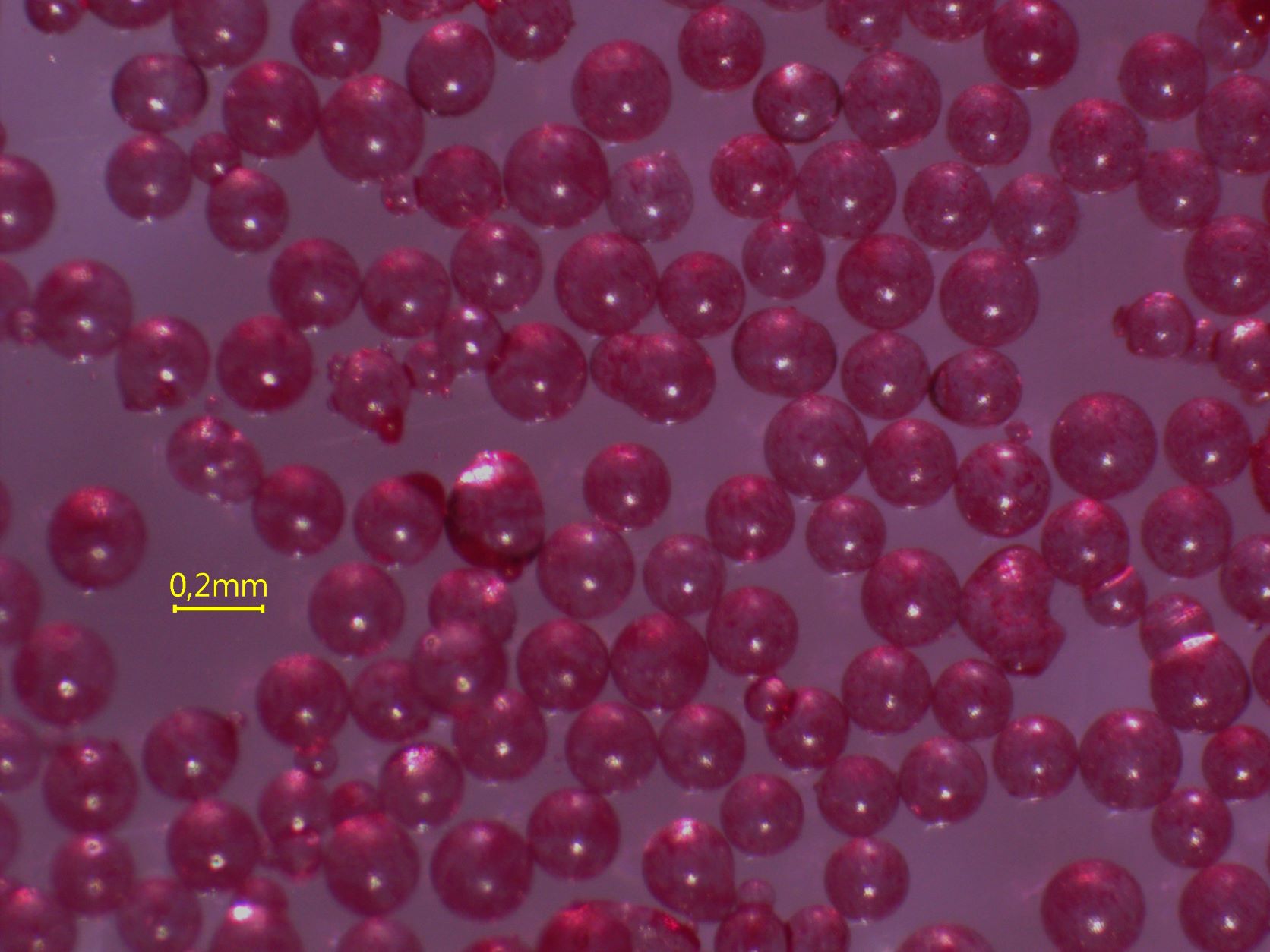}
	\end{center}
	\caption{Microscopy image for the 0.15 mm $\leq$ $d$ $\leq$ 0.25 mm round glass beads of red color (species 1).}
	\label{fig3}
\end{figure}

\begin{figure}[b!]
	\begin{center}
		\includegraphics[width=0.7\linewidth]{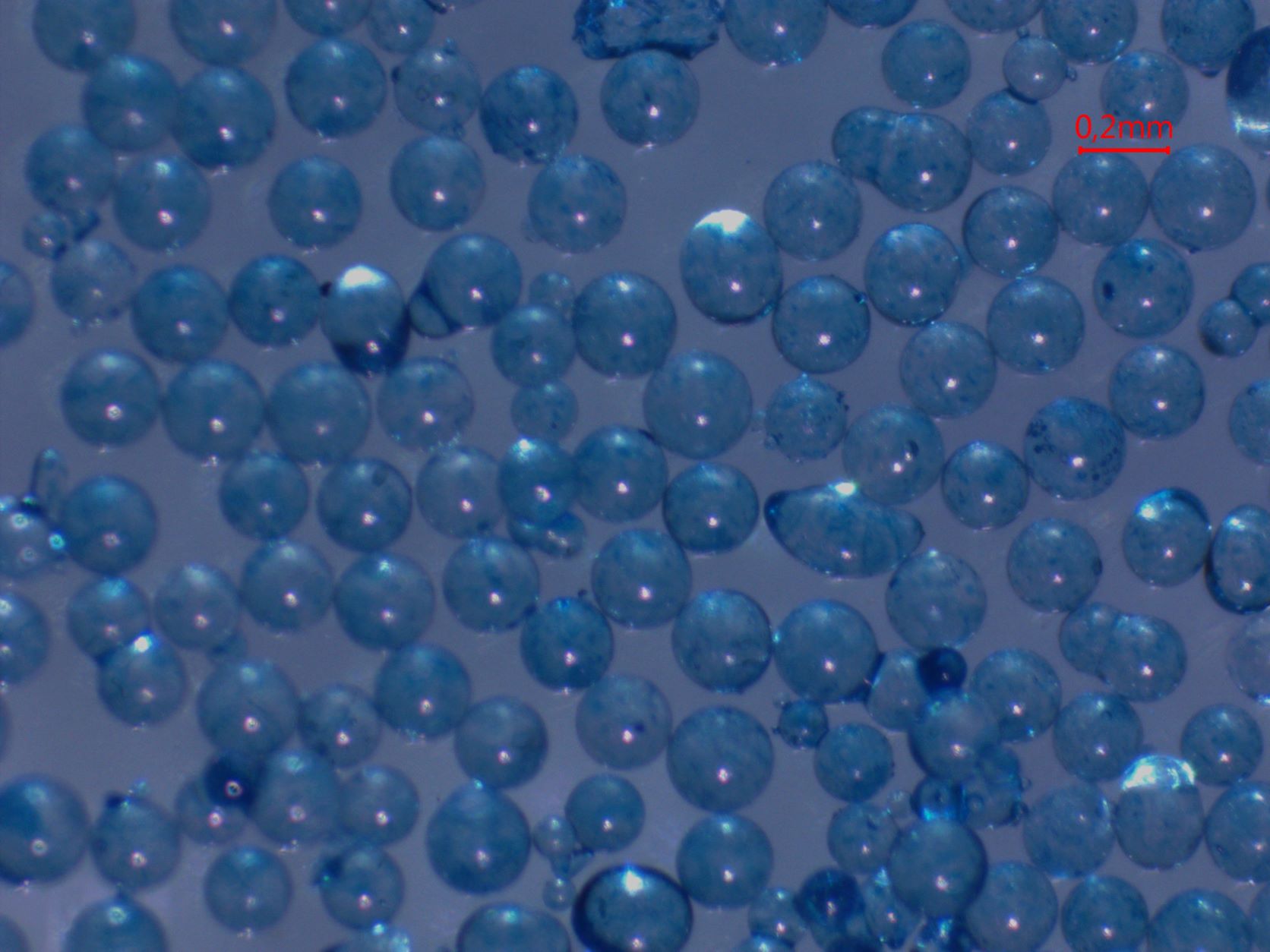}
	\end{center}
	\caption{Microscopy image for the 0.15 mm $\leq$ $d$ $\leq$ 0.25 mm round glass beads of blue color (species 1).}
	\label{fig4}
\end{figure}

\begin{figure}[h!]
	\begin{center}
		\begin{minipage}{0.49\linewidth}
			\begin{tabular}{c}
				\includegraphics[width=0.99\linewidth]{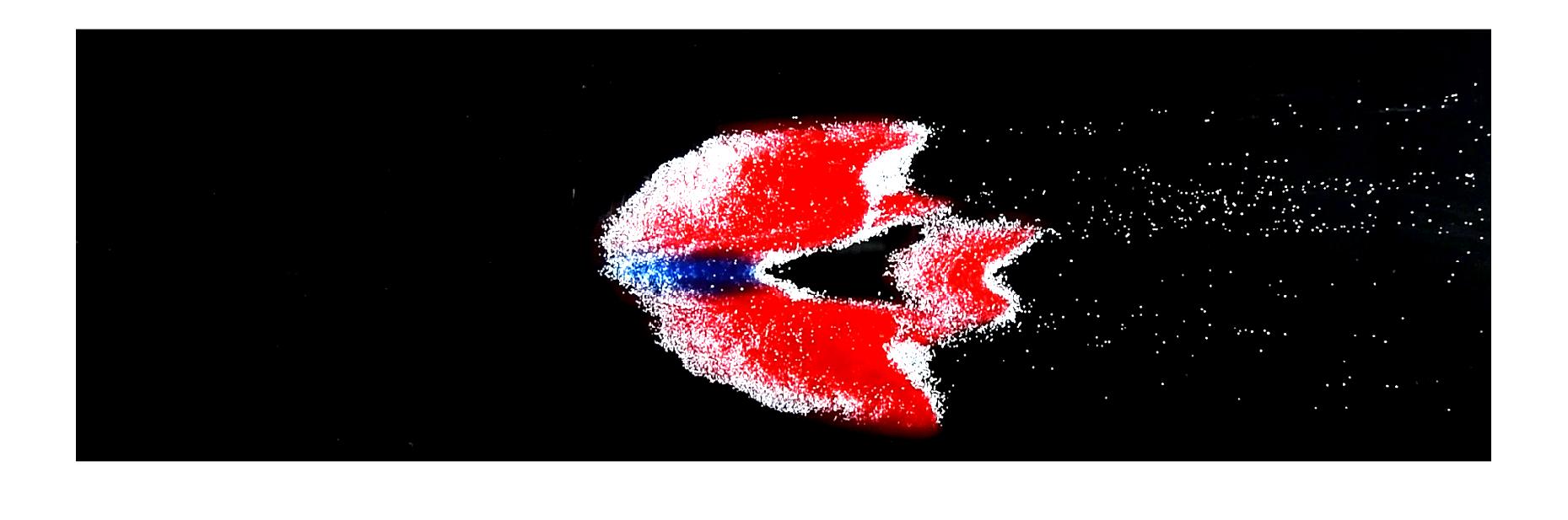}\\
				(a)
			\end{tabular}
		\end{minipage}
		\hfill
		\begin{minipage}{0.49\linewidth}
			\begin{tabular}{c}
				\includegraphics[width=0.99\linewidth]{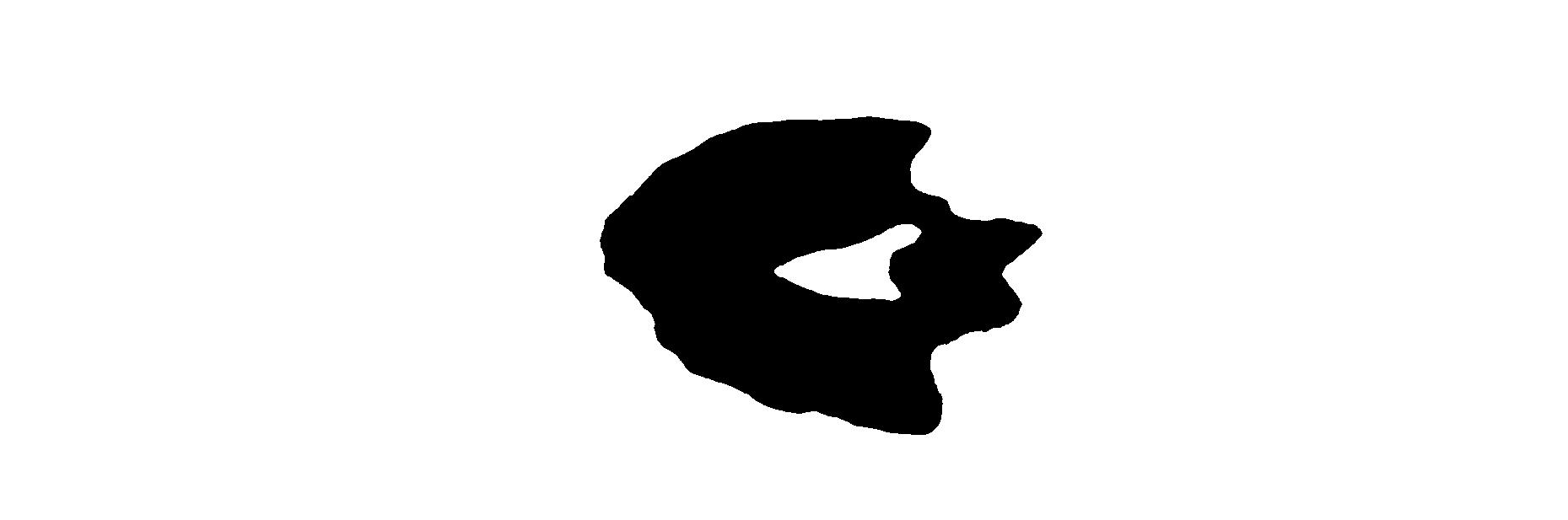}\\
				(b)
			\end{tabular}
		\end{minipage}
		\hfill
		\begin{minipage}{0.49\linewidth}
			\begin{tabular}{c}
				\\
				\includegraphics[width=0.99\linewidth]{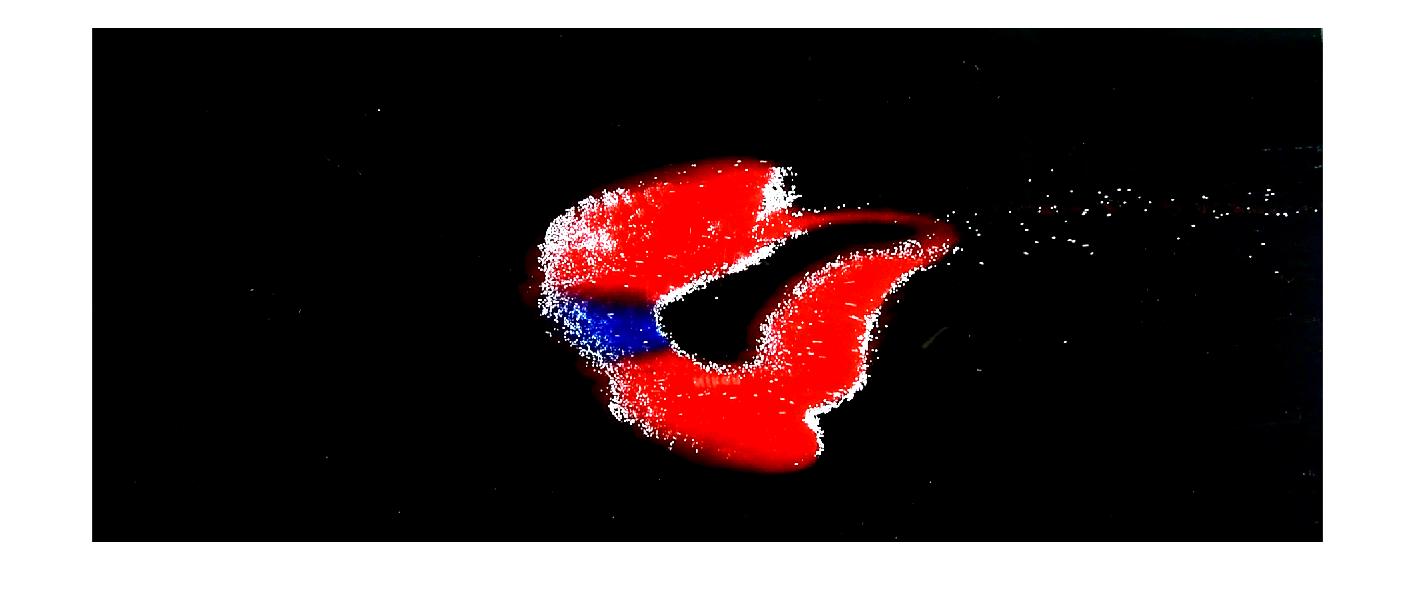}\\
				(c)
			\end{tabular}
		\end{minipage}
		\hfill
		\begin{minipage}{0.49\linewidth}
			\begin{tabular}{c}
				\\
				\includegraphics[width=0.99\linewidth]{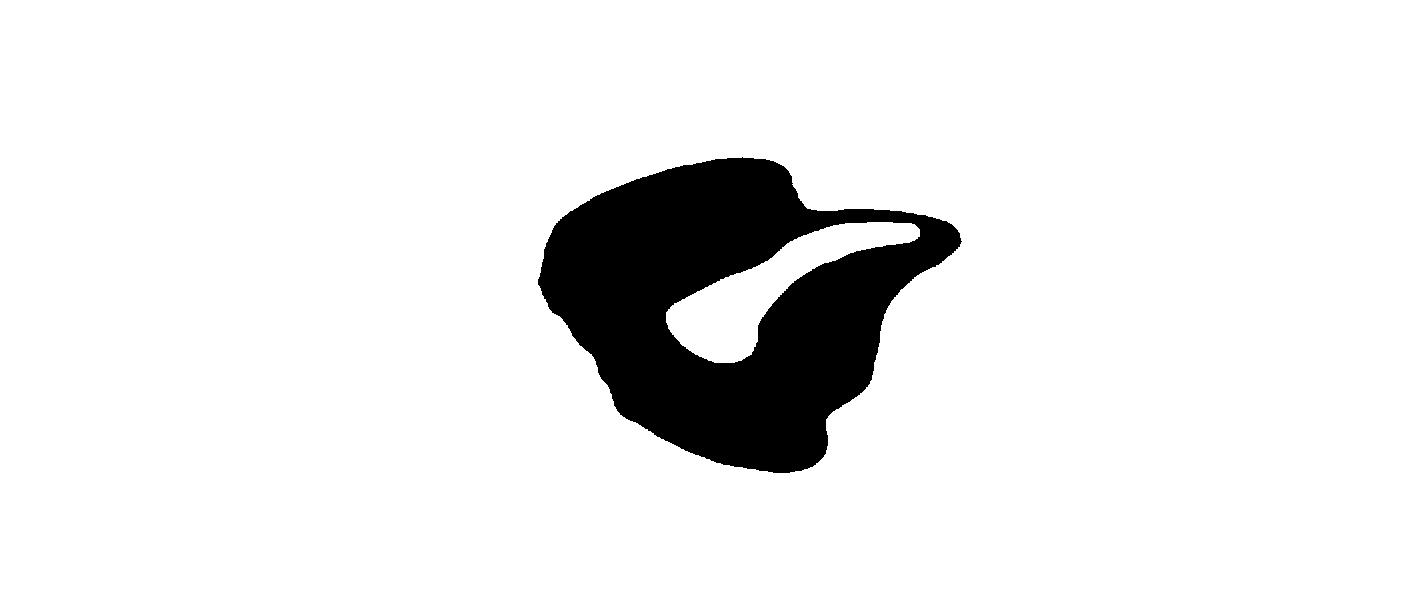}\\
				(d)
			\end{tabular}
		\end{minipage}
		\hfill
	\end{center}
	\caption{Void regions during the exchange processes for (a) and (b) $\phi_1$ = $\phi_2$ = 0.5 and (c) and (d) $\phi_1$ $\neq$ $\phi_2$. Figures (a) and (c) are raw images and figures (b) and (d) binarized images.}
\end{figure}

\begin{figure}[b!]
	\begin{center}
		\includegraphics[width=0.5\linewidth]{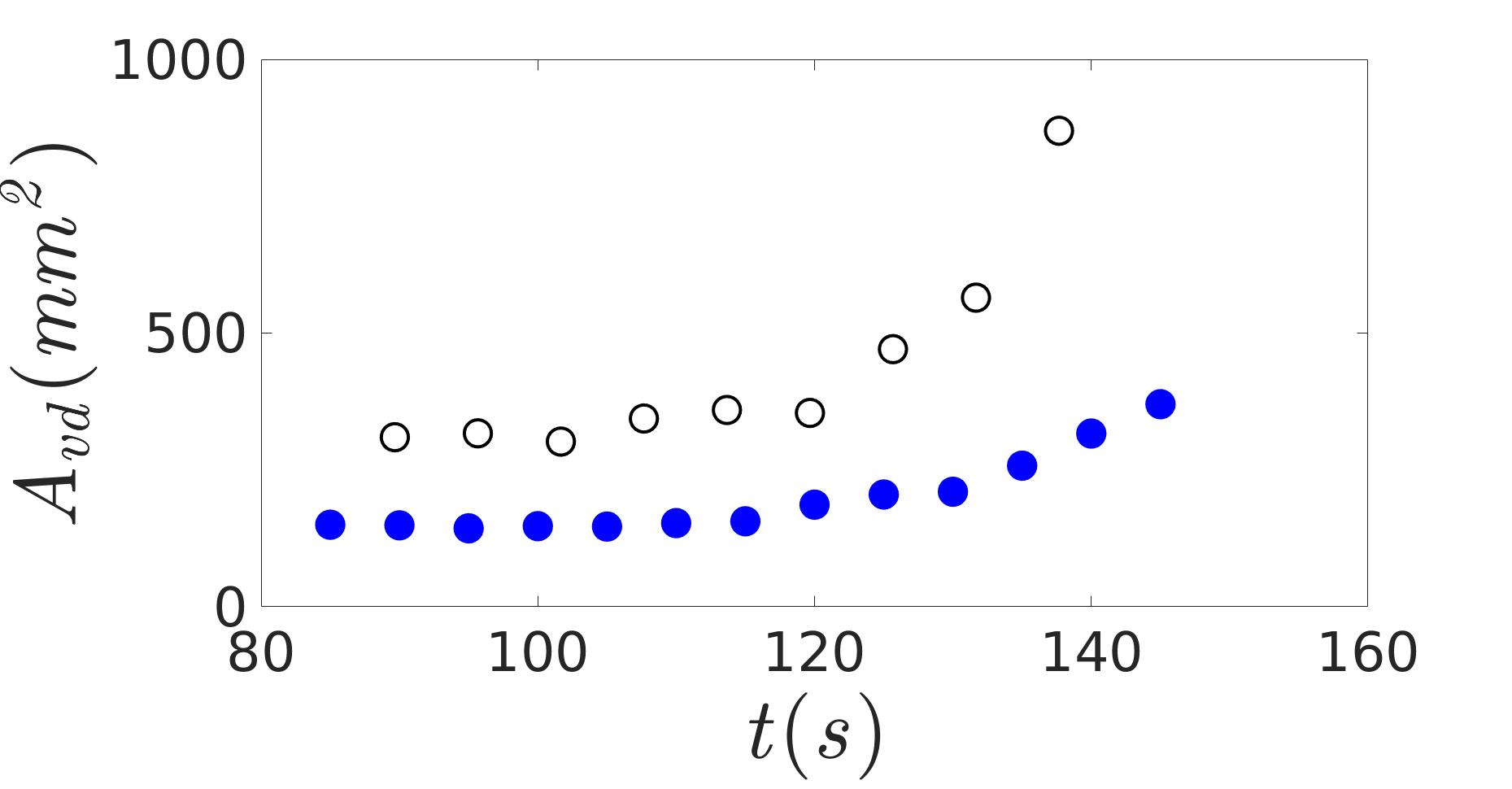}
	\end{center}
	\caption{Area occupied by the void region $A_{vd}$ as a function of time, in dimensional form, for the case of bidisperse piles (mixtures). Solid blue circles correspond to the exchange pattern when $\phi_1$ = $\phi_2$ = 0.5 and open black symbols when $\phi_1$ $\neq$ $\phi_2$.}
\end{figure}

\begin{figure}[h!]
	\begin{center}
		\includegraphics[width=.5\linewidth]{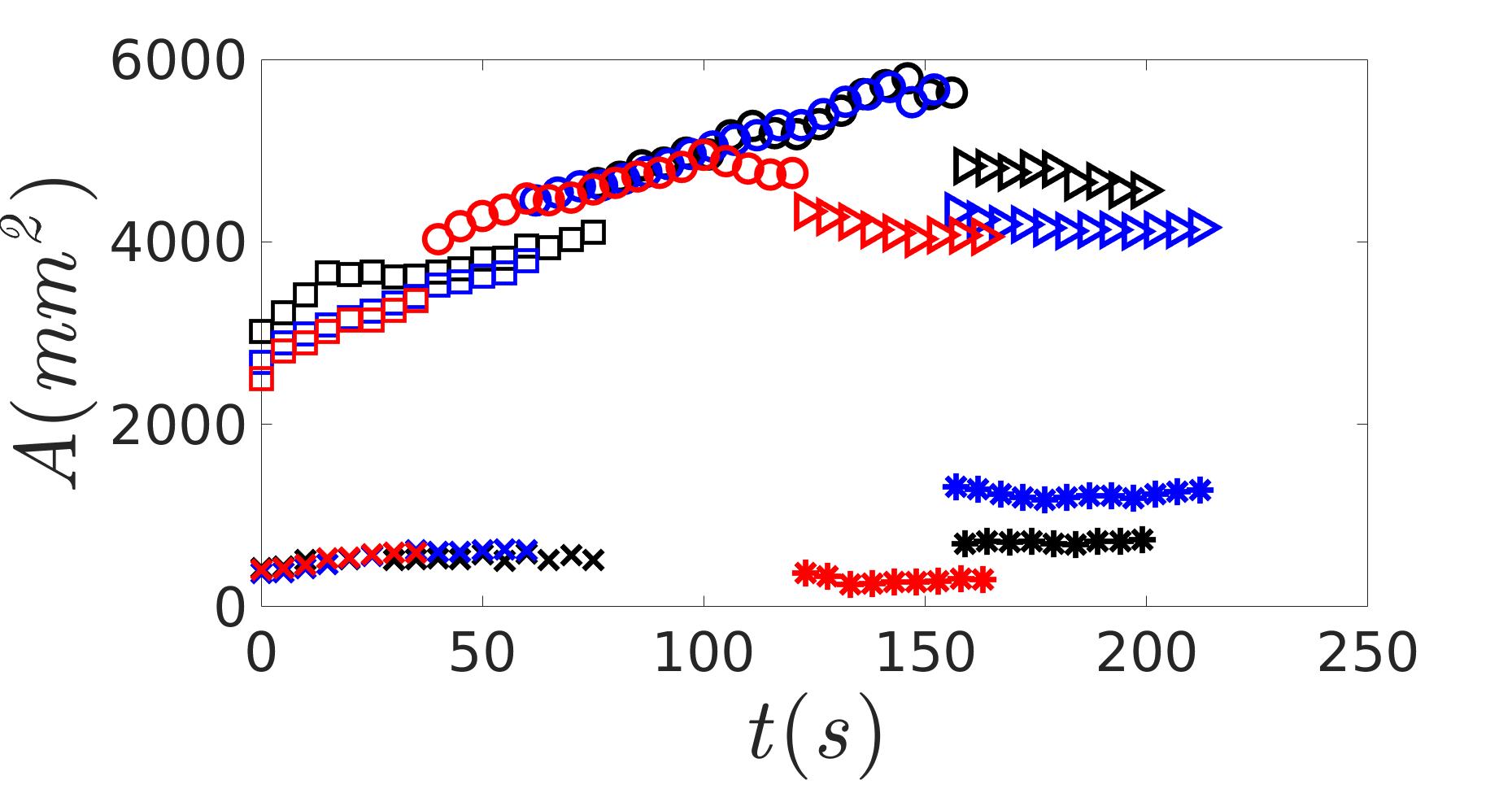}
	\end{center}
	\caption{Projected areas, in dimensional form, of impact (x), target ($\Box$), merged ($\circ$), parent ($\triangleright$) and baby barchans (*), respectively, as functions of time, for exchange processes with bidisperse piles (mixtures). Black, blue and red colors correspond to cases $a$, $f$ and $h$.}
\end{figure}

\end{document}